\documentclass[referee, useAMS,usenatbib,usegraphicx]{mn2e}

\title[Polarimetric study towards NGC 1893]{Multi-wavelength polarimetric study towards the open cluster NGC 1893}
\author[C. Eswaraiah et al.]{C. Eswaraiah,$^{1}$\thanks{E-mail:eswar@aries.res.in} 
A. K. Pandey,$^{1}$ G. Maheswar,$^{1}$ Biman J. Medhi,$^{1}$  
\newauthor{J. C. Pandey$^{1}$, D. K. Ojha$^{2}$ and W. P. Chen$^{3}$} \\ 
$^{1}$Aryabhatta Research Institute of Observational Sciences, Manora Peak, Nainital 263129, India\\
$^{2}$Tata Institute of Fundamental Research, Mumbai 400 005, India\\
$^{3}$Institute of Astronomy, National Central University, Chung-Li 32054, Taiwan}
\begin{document}

\date{Accepted ------------, Received ------------; in original form ------------}

\pagerange{\pageref{firstpage}--\pageref{lastpage}} \pubyear{2010}

\maketitle

\label{firstpage}

\begin{abstract}
We present multi-wavelength linear polarimetric observations 
for 44 stars of the NGC 1893 young open cluster region along with 
$V$-band polarimetric observations of stars of other four 
open clusters located between $l$ $\sim$ 160$\degr$ to 175$\degr$. 
We found evidence for the presence of two dust layers located at a distance of 
$\sim$ 170 pc and $\sim$ 360 pc. 
The dust layers produce a polarization $P_{V}$ $\sim$ $2.2\%$. 
It is evident from the clusters studied in the present work that, 
in the Galactic longitude range 
$l$ $\sim$ $160\degr$ to $175\degr$ and within the 
Galactic plane ($|b|\textless 2\degr$), 
the polarization angles remain almost constant, 
with a mean $\sim$ 163$\degr$ and a 
dispersion of 6$\degr$. The small dispersion in polarization angle 
could be due to the presence of 
uniform dust layer beyond 1 kpc. 
Present observations reveal that in case of NGC 1893, 
the foreground two dust layers, in addition to the intracluster medium, 
seems to be responsible for the polarization effects. 
It is also found that towards the direction of NGC 1893, 
the dust layer that exists between 2-3 kpc has a negligible contribution 
towards the total observed polarization. 
The weighted mean for percentage of polarization ($P_{max}$) and
the wavelength at maximum polarization ($\lambda_{max}$) are found to be
$2.59 \pm 0.02\%$ and  $0.55 \pm 0.01$ $\mu$m respectively.
The estimated mean value of $\lambda_{max}$ indicates that
the average size of the dust grains within 
the cluster is similar to that 
in the general interstellar medium.
The spatial variation of the polarization is found to decrease 
towards the outer region of the cluster. 
In the present work, we support the notion, as already has been 
shown in previous studies, that polarimetry, 
in combination with $(U-B)/(B-V)$ colour-colour diagram, 
is a useful tool for identifying 
non-members in a cluster. 

\end{abstract}

\section{INTRODUCTION}

Dust properties govern several physical and chemical phenomena 
in the interstellar medium (ISM) and act as a tracer of local environmental conditions. 
When starlight passes through various components of interstellar dust grains, 
which are aligned to the Galactic magnetic field, 
the radiation becomes partially plane-polarized, 
typically at the level of a few percent. 
The nature of this polarization 
reveals important information
regarding the shape, size and composition 
of interstellar dust. The detailed process(es) by which grains get aligned 
with the magnetic fields has (have) long been, 
and is still actively under study 
(Davis \& Greenstein 1951; Dolginov 1990; Cho \& Lazarian 2005). 
In general the dust grains tend to align with their long axis 
perpendicular to the magnetic field (Purcell 1979). 
The clue to the interstellar origin of the polarization came from the observed 
correlation between the degree of polarization ($P$) and the colour excess, $E(B-V)$. 
The values of $P$ (in visual wavelengths) for the stars with large $E(B-V)$ 
are found to be in the range between zero and a maximum value given by 
$P_{max}/E(B-V)=~9\%~mag^{-1}$ (Aannestad \& Purcell 1973). 
The relation between $P_{max}$ and the colour excess, and the variation of
$P$ with wavelength, are interpreted in terms of the grain properties and
the efficiency of the grain alignment. 
Therefore, polarimetry is a useful technique to 
investigate the properties like maximum polarization $P_{\lambda_{max}}$, 
the wavelength $\lambda_{max}$ corresponding to $P_{\lambda_{max}}$ 
and the orientation of the magnetic field in various Galactic locations.

The photometric and spectroscopic information already available for stars 
in open clusters are specifically important to make a meaningful study of the 
dust grains located in the foreground and in the intracluster regions 
(e.g., Trumpler 27, Feinstein et al. 2000; NGC 6231, Feinstein et al. 2003; 
Stock 16, Feinstein et al. 2003; 
Hogg 22 \& NGC 6204, Mart\'{i}nez et al. 2004; 
NGC 5606, Orsatti et al. 2007; NGC 5749, Vergne et al. 2007; 
IC 1805, Medhi et al. 2007; NGC 6250, Feinstein et al. 2008; 
NGC 654, Medhi et al. 2008; NGC 6124, Vergne et al. 2010; 
NGC 6823, Medhi et al. 2010). In this paper, we present multi-wavelength 
polarimetry of stars towards NGC 1893. We also present 
$V$-band polarimetry of stars towards 
four additional clusters, $\it{viz}$ NGC 2281, NGC 1664, NGC 1960 and Stock 8, 
to get information about the ISM foreground to the cluster NGC 1893. 
The main aim of this 
study is to investigate the dust properties as a function of distance towards 
the anticenter direction of the Galaxy ($l$ $\sim160\degr$ to $175\degr$) 
using stars in open clusters.  

The basic parameters of the observed clusters are given in Table \ref{basic_para}. 
Using these five open clusters, we have made an attempt to 
study the dust properties of the ISM distributed between 
$\sim600$ pc (NGC 2281) and $\sim3$ kpc (NGC 1893). The paper is organized 
in the following manner. In section \ref{instu_obs_data}, 
we present a brief discussion on the observations and data reduction. 
We present our results in section \ref{results}, and analysis and discussion in section \ref{discuss}. 
The dust components responsible for the observed polarization are discussed in section \ref{dust_components}. 
We conclude our results 
in section \ref{conclude}.
\begin{table*}
\centering
\caption{The basic parameters of the clusters.}\label{basic_para}
\begin{tabular}{lcccccc}\hline \hline
Cluster Id & l       &  b      &  Distance & $E(B-V)$ & log(age) & Reference  \\
           & $(\degr)$& $(\degr)$ & (pc)  &  (\it{mag})  & ({\it yr})  &      \\
\hline
NGC 2281  &174.90 &$+$16.88  & 558   & 0.06, 0.11  & 8.70 & Karchenko et al. (2005), Glaspey (1987) \\
NGC 1664  &161.68 &$-$0.45   & 1199  & 0.25  & 8.72 & Karchenko et al. (2005) \\
NGC 1960  &174.54 &$+$1.07   & 1330  & 0.22  & 7.40 & Sharma et al. (2006)    \\
Stock 8   &173.37 &$-$0.18   & 2050  & 0.40$-$0.60 & 6.00$-$6.70 & Jose et al. (2008) \\
NGC 1893  &173.59 &$-$1.68   & 3250  & 0.40$-$0.60 & 6.60 & Sharma et al. (2007)  \\
\hline
\end{tabular}
\end{table*}
\begin{table*}
\centering
\caption{The observational details}\label{obs_log}
\begin{tabular}{lcc}\hline \hline
Cluster Id & Date of Observation  &  Passband(s)  \\
           &  (year, month, date)  &            \\
\hline
NGC 2281   &  2009, 12, 24       & \it{V} \\
NGC 1664   &  2009, 12, 24       & \it{V} \\
NGC 1960   &  2009, 11, 23       & \it{V} \\
Stock 8    &  2009, 11, 23       & \it{V} \\
NGC 1893   &  2008, 11, 8 $\&$ 9 & \it{BV$(RI)_{c}$} \\
\hline
\end{tabular}
\end{table*}

\section{Observations and Data Reduction}\label{instu_obs_data}
Polarimetric observations were carried out 
using the ARIES Imaging Polarimeter 
(AIMPOL: Rautela, Joshi \& Pandey, 2004; Medhi et al. 2007, 2010) 
mounted at the Cassegrain 
focus of the 104-cm Sampurnanand telescope of the Aryabhatta 
Research Institute of observational sciencES (ARIES), Nainital, India, 
coupled with a TK 1024$\times$1024 pixel$^2$ CCD camera. 
The AIMPOL consists of a half-wave plate (HWP) modulator and a Wollaston prism beam-splitter. 
The observations were carried out in $B$, $V$, $R_{c}$ and $I_{c}$  
($\lambda_{B_{eff}}$=0.440 $\mu$m, $\lambda_{V_{eff}}$=0.530 $\mu$m,  
$\lambda_{Rc_{eff}}$=0.670 $\mu$m and $\lambda_{I_{eff}}$=0.800 $\mu$m) 
photometric bands. 
Details of the observations are given in Table \ref{obs_log}.
Each pixel of the CCD corresponds to 1.73 arcsec and the 
field-of-view is $\sim$~8 arcmin in diameter. The FWHM of the stellar image 
varied from 2 to 3 pixels. The read out noise and gain of the CCD are 7.0 $e^{-1}$  
and 11.98 $e^{-1}$/ADU, respectively. Due to the absence of a grid in AIMPOL, 
we manually checked for any overlap of ordinary and extra-ordinary images of the sources.

Fluxes of ordinary ($I_{o}$) and extra-ordinary ($I_{e}$) beams for all the observed sources 
with good signal-to-noise ratio were extracted by standard aperture photometry 
after bias subtraction using the {\small IRAF}\footnote{{\small IRAF} is distributed by National Optical 
Astronomical Observatories, USA.} package. 
The ratio $R(\alpha)$ 
is given by: 
\begin{equation}\label{R_alpha}
 R(\alpha) = \frac{\frac{{I_{e}}(\alpha)}{{I_{o}}(\alpha)}-1} {\frac{I_{e}(\alpha)} {I_{o}(\alpha)}+1} =  P \cos(2\theta - 4\alpha),
\end{equation}
where $P$ is the fraction of the total linearly polarized light and $\theta$
is the polarization angle of the plane of polarization. 
Here $\alpha$ is the position of the fast axis of the HWP at
0$\degr$, 22.5$\degr$, 45$\degr$ and 67.5$\degr$ corresponding to the four
normalized Stokes parameters respectively, $q$ [R(0$\degr$)], $u$ [R(22.5$\degr$)], 
$q_{1}$ [R(45$\degr$)] and $u_{1}$ [R(67.5$\degr$)]. 
The detailed procedures used to estimate the polarization and polarization 
angles for the programme stars are described by Ramaprakash et al. (1998), 
Rautela, Pandey, \& Joshi (2004) and Medhi et al. (2010).
Since polarization accuracy is, in principle, limited by photon statistics,
we estimated the errors in normalized Stokes parameters
$\sigma_{R(\alpha)}$ ($\sigma_{q}$, $\sigma_{u}$, $\sigma_{q_{1}}$ and
$\sigma_{u_{1}}$ in $\%$) using the expression (Ramaprakash et al. 1998):
\begin{equation}\label{stok_err}
\sigma_{R(\alpha)}=\sqrt{(N_{e}+N_{o}+2N_{b})}/(N_{e}+N_{o}) \\
\end{equation} 
where $N_{e}$ and $N_{o}$ are the counts in extra-ordinary and ordinary rays 
respectively, and $N_{b}[=\frac{N_{be}+N_{bo}}{2}]$ is the average background counts 
around the extra-ordinary and ordinary rays of a source. 
The individual errors associated with the four values of $R(\alpha)$,
estimated using equation (\ref{stok_err}), are used      
as weights in calculation of $P$ and $\theta$ for the programme stars.

To correct the measurements for null polarization 
(or instrumental polarization) and
the zero-point polarization angle, we observed several 
polarized and unpolarized standards
taken from Schmidt, Elston \& Lupie (1992).
The results on polarized and unpolarized standards are given in Table \ref{tab_standards}.
The values of $\theta$ are in equatorial coordinate system measured from the
North increasing towards the East. Both the observed degree of polarization [$P(\%)$]
and polarization angle [$\theta (\degr)$] for 
the polarized standards are
in good agreement with those given by Schmidt et al. 1992. 
The observed normalized Stokes parameters $q$ and $u$ 
(q\%, u\%) for standard unpolarized stars are also given in Table \ref{tab_standards}.
The instrumental polarization of AIMPOL on the 104-cm Sampurnanand Telescope has been 
monitored since 2004 in different projects and found to be 
less than 0.1$\%$ in different bands (Rautela, Pandey, \& Joshi, 2004; Medhi et al. 2007, 2008, 2010; 
Pandey et al. 2009).
\begin{figure*}
\resizebox{15cm}{15cm}{\includegraphics{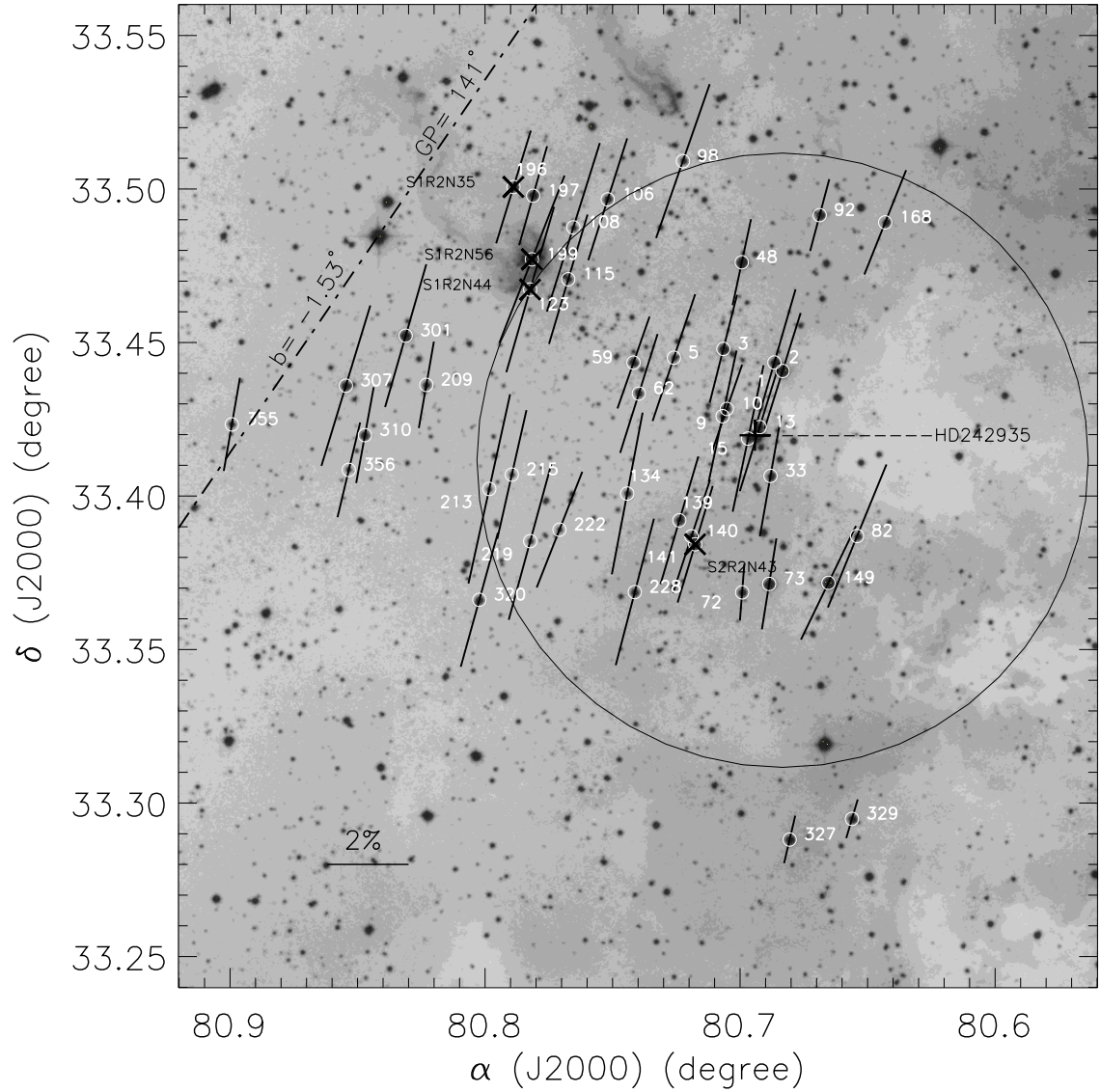}}
\caption{Stellar polarization superimposed on 
the 20$^{\arcmin} \times$ 20$^{\arcmin}$ ${\it{R}}$-band DSS II 
image of the field containing NGC 1893. 
The length of each polarization vector is
proportional to $P_{V}$. A vector with a polarization of 2$\%$ is drawn for 
reference.  The dash-dotted line marks the Galactic parallel $b$=-1.53$\degr$. 
H$\alpha$ emission stars identified by Marco and Negueruela (2002)
(S2R2N43) and by Negueruela et al (2007) (S1R2N35, S1R2N56 and S1R2N44) are
shown with cross symbols.
The identification numbers are taken from Cuffey and Shapley (1937).
The circle represents the boundary ($\sim 6^{\arcmin}$ radius, Sharma et al. 2007) of the cluster.} 
\label{NGC1893V_band}
\end{figure*}

\section{Results}\label{results}
\subsection{NGC 1893}

The polarization measurements for stars in the region of NGC 1893 
are listed in Table \ref{result_1893}. 
The star identification numbers (column 1)
are taken from Cuffey and Shapley (1937). The $V$-band magnitudes are given in column 2.
The degree of polarization $P$ (in percent) 
and polarization angles $\theta$ (in degree) measured in $B$, $V$, $(R,I)_{c}$ bands 
and their corresponding standard errors ($\epsilon_{P}$ and $\epsilon_{\theta}$) are given 
in columns 3 to 10. 

The sky projection of the $V$-band polarization vectors for the 44 stars 
measured towards the NGC 1893 region is drawn on the ${\it{R}}$-band Digitized Sky Survey II (DSS II) 
image (Fig. \ref{NGC1893V_band}). 
The length of each polarization vector is proportional to the degree of polarization. 
A vector with a polarization of 2$\%$ is drawn for reference. 
The dash-dotted line superimposed 
in Fig. \ref{NGC1893V_band} is 
indicating the orientation of the projection of the Galactic plane (GP) 
which has a polarization angle of $\theta_{GP}=141\degr$.
The average polarization ($P$) and polarization angle ($\theta$) 
for 44 stars are found to be $2.4 \pm 0.6\%$, $160 \pm 5\degr$ in $B$, $2.6 \pm 0.7\%$, 
$162 \pm 5\degr$ in $V$, $2.5 \pm 0.7\%$, $159 \pm 6\degr$ in $R_{c}$ 
and $2.2 \pm 0.5\%$, $161 \pm 6\degr$ in $I_{c}$ bands, respectively. 

\subsection{NGC 2281, NGC 1664, NGC 1960 and Stock 8}
Polarimetric measurements in $V$-band for 
14 stars towards NGC 2281, 27 stars towards NGC 1664, 15 stars towards NGC 1960 and 
21 stars towards Stock 8, have also been carried out.
The results are listed in 
Tables \ref{result_2281}, \ref{result_1664}, \ref{result_1960} and \ref{result_stock8} 
respectively. The star identification numbers 
for NGC 2281, NGC 1664, NGC 1960 and Stock 8 
are taken from Vasileviskis \& Balz (1959), Larsson-Leander (1957), Boden (1951) and Mayer (1964) 
respectively. 

The mean values of $P_{V}$ and $\theta_{V}$ 
towards NGC 2281, NGC 1664, NGC 1960 and Stock 8 are found to be 
$0.9\pm0.2\%$ and $16\pm6\degr$; $1.8\pm0.7\%$ and $172\pm8\degr$; 
$1.2\pm0.2\%$ and $159\pm5\degr$ and $2.4\pm0.6\%$ \& $160\pm7\degr$, 
respectively. 

\begin{figure*}
\resizebox{18cm}{17cm}{\includegraphics{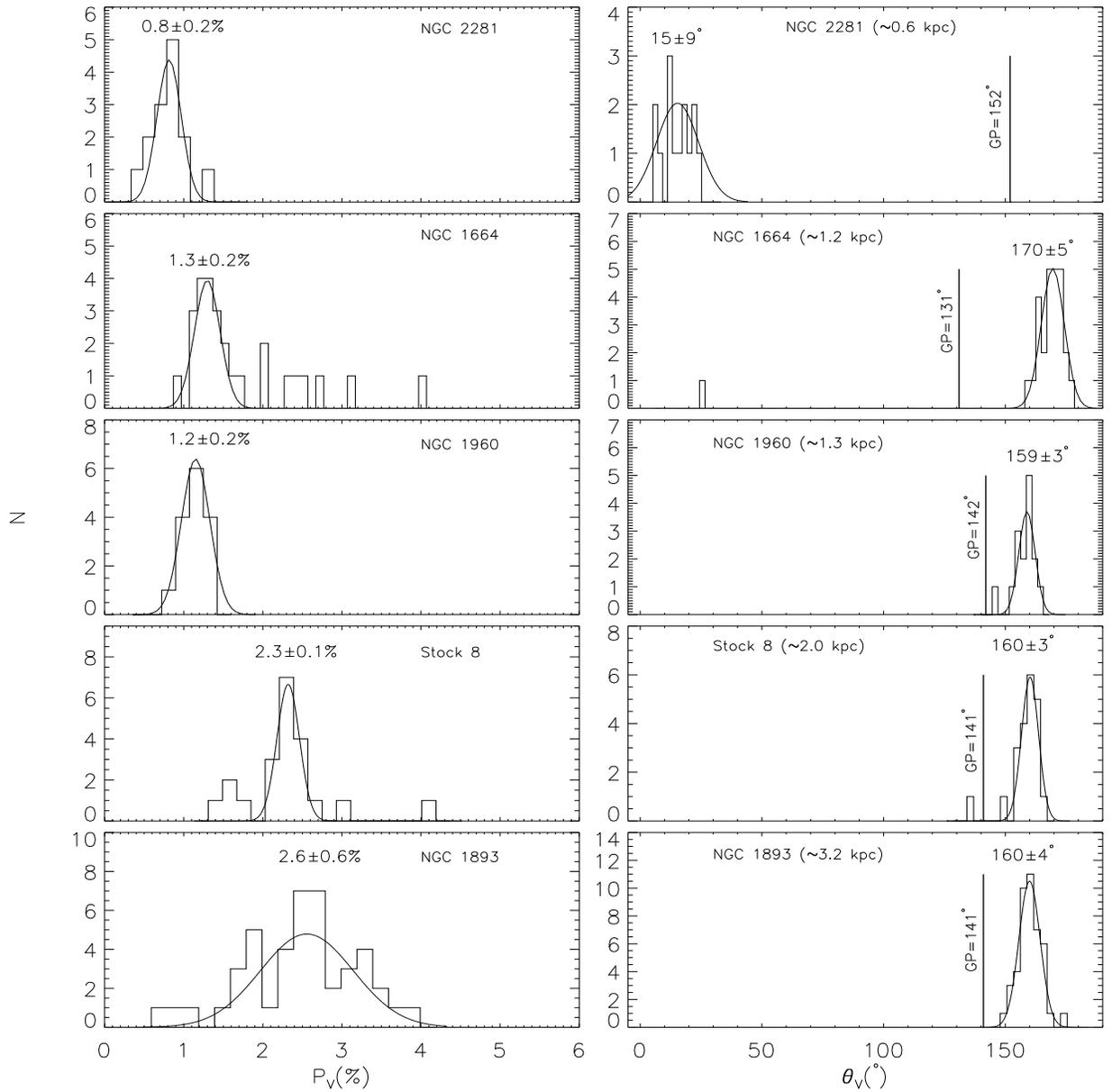}}
\caption{Histograms for $P_{V}$ (left panel) and $\theta_{V}$ (right panel) 
in five observed clusters, NGC 2281, NGC 1664, NGC 1960, 
Stock 8 and NGC 1893. 
The polarization angle for the Galactic plane of 
each cluster is shown with a continuous line and is labeled with GP. 
Each histogram is fitted with a Gaussian function.}
\label{Hist_5}
\end{figure*}

\section{Analysis and Discussion}\label{discuss}
Fig. \ref{Hist_5} presents the distribution of $P_{V}$ and $\theta_{V}$ 
for stars in the observed cluster regions. 
Four clusters studied here, namely NGC 1664, NGC 1960, Stock 8 and NGC 1893 
are located at Galactic latitude $|b|<2\degr$. 
The nearest open cluster NGC 2281 studied here is located at $b=+16.88\degr$. 
A Gaussian fit to the distribution of $P_{V}$ and $\theta_{V}$ yields a mean and a standard deviation 
as $0.8\pm0.2\%$ and $15\pm9\degr$ for NGC 2281; $1.3\pm0.2\%$ and $170\pm5\degr$ 
for NGC 1664; $1.2\pm0.2\%$ and $159\pm3\degr$ for NGC 1960; 
$2.3\pm0.1\%$ and $160\pm3\degr$ for Stock 8; and $2.6\pm0.6\%$ and $160\pm4\degr$ for NGC 1893. 
It is evident from Fig. \ref{Hist_5} that the mean value of $P_{V}$ 
increase with the distance of a cluster. 

The polarization angle corresponding to the Galactic plane 
for each cluster is drawn with a continuous line. Fig. \ref{Hist_5} (right panel) 
indicates that the difference between GP and mean $\theta_{V}$ is the largest 
for the nearby least polarized ($0.8\pm0.2\%$) cluster 
(NGC 2281: $l$=174.90$\degr$ and $b$=$+$16.88$\degr$). 
Whereas, in the case of four clusters, namely NGC 1664, NGC 1960, Stock 8 and NGC 1893, 
the difference between GP and mean $\theta_{V}$ (Gaussian) is small and is almost the same. 
Fig. \ref{Hist_5} also reveals that in the Galactic 
longitude range $l\sim160{\degr}$ to $\sim175{\degr}$ and 
within the Galactic plane ($|b|<2\degr$), the polarization angles remain almost constant. 
The average of the mean values of polarization angles for the 
four clusters is found to be $163\degr$ with a standard deviation of $6\degr$. 
This implies that at least beyond $\sim$1 kpc the magnetic field orientation 
remains almost unchanged. However, for a smaller distance ($\sim 0.6$ kpc), 
the polarization angle towards the northern hemisphere (NGC 2281) of the 
Galactic plane is found to be $16\degr$ with a standard deviation 
of $6\degr$. 

The polarization towards the Taurus molecular complex 
($l=174.13\degr$, $b=-13.45\degr$), taken from Heiles (2000), reveals 
a mean polarization angle of $58\degr$ with a standard deviation of $38\degr$. 
The region selected around the Taurus molecular 
complex is bound between the Galactic longitude $l=160\degr$ to $175\degr$ and latitude $b<-2\degr$.
It is interesting to mention that, the mean polarization angles at high Galactic latitude 
($|b|>2\degr$) are significantly different from those in the Galactic plane $|b|<2\degr$. 
\begin{figure*}
\resizebox{15.5cm}{19.5cm}{\includegraphics{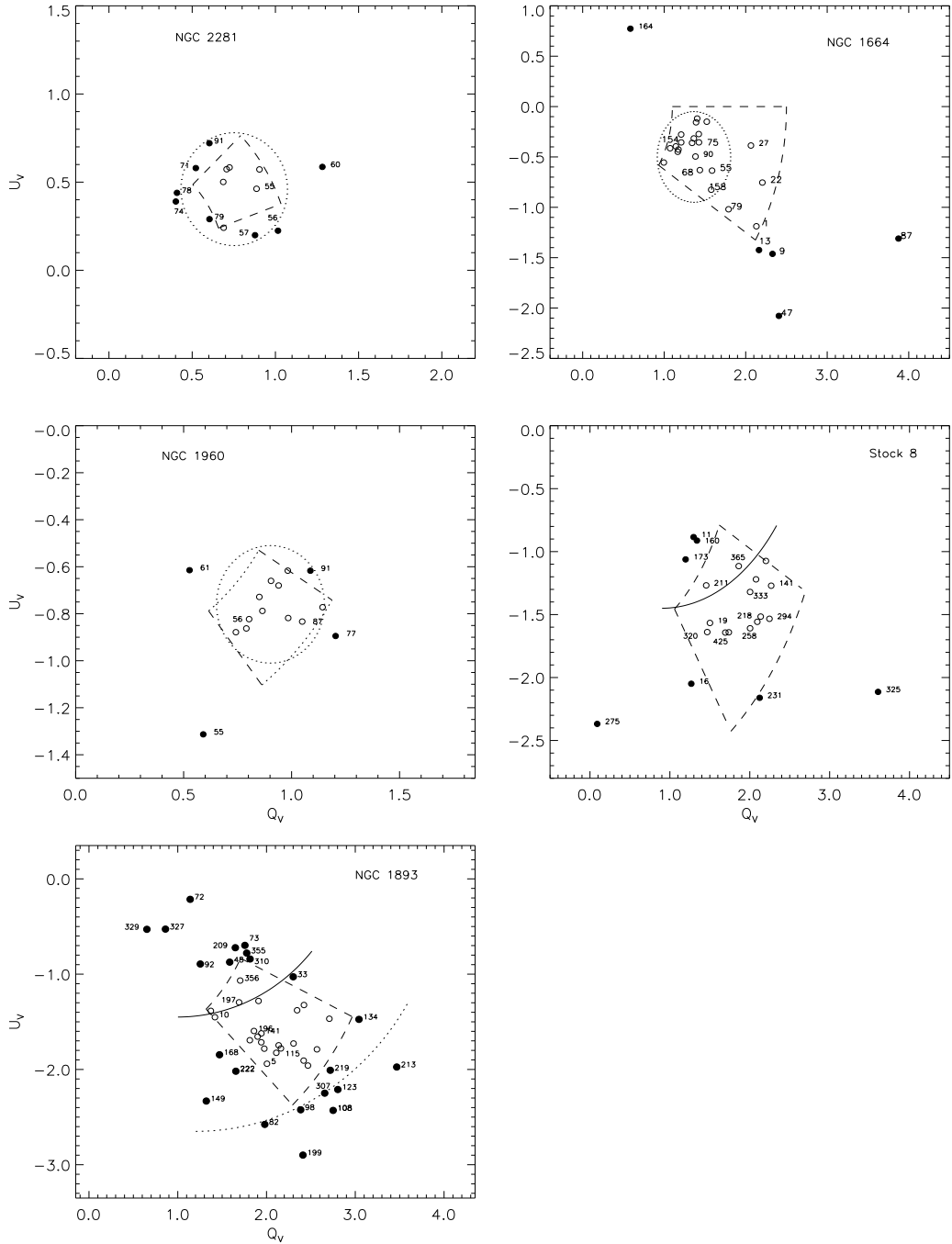}}
\caption{The $U_{V}$ versus $Q_{V}$ diagrams for stars towards 
NGC 2281, NGC 1664, NGC 1960, Stock 8 and NGC 1893 cluster regions, measured in the $V$-band. 
In each case a box with dashed line 
marks the boundary of mean $P_{V} \pm \sigma$ and mean $\theta_{V} \pm \sigma$. 
Stars lying outside the 1$\sigma$ box are shown by 
filled circles. It can be seen that the majority of the stars of apparent 
grouping lie within the box (open circles). 
The apparent grouping in the case of NGC 2281, NGC 1664 and NGC 1960 
is bound by a dotted circle. The thin continuous curve in the case of Stock 8 and NGC 1893 separates 
foreground field stars from cluster members. The dotted curve in the case of NGC 1893 demarcates the 
stars having relatively high polarization and high $E(B-V)$ values as well as having intrinsic 
polarization (see the text).} \label{U_Q4}
\end{figure*}
\begin{figure*}
\resizebox{17.5cm}{17.5cm}{\includegraphics{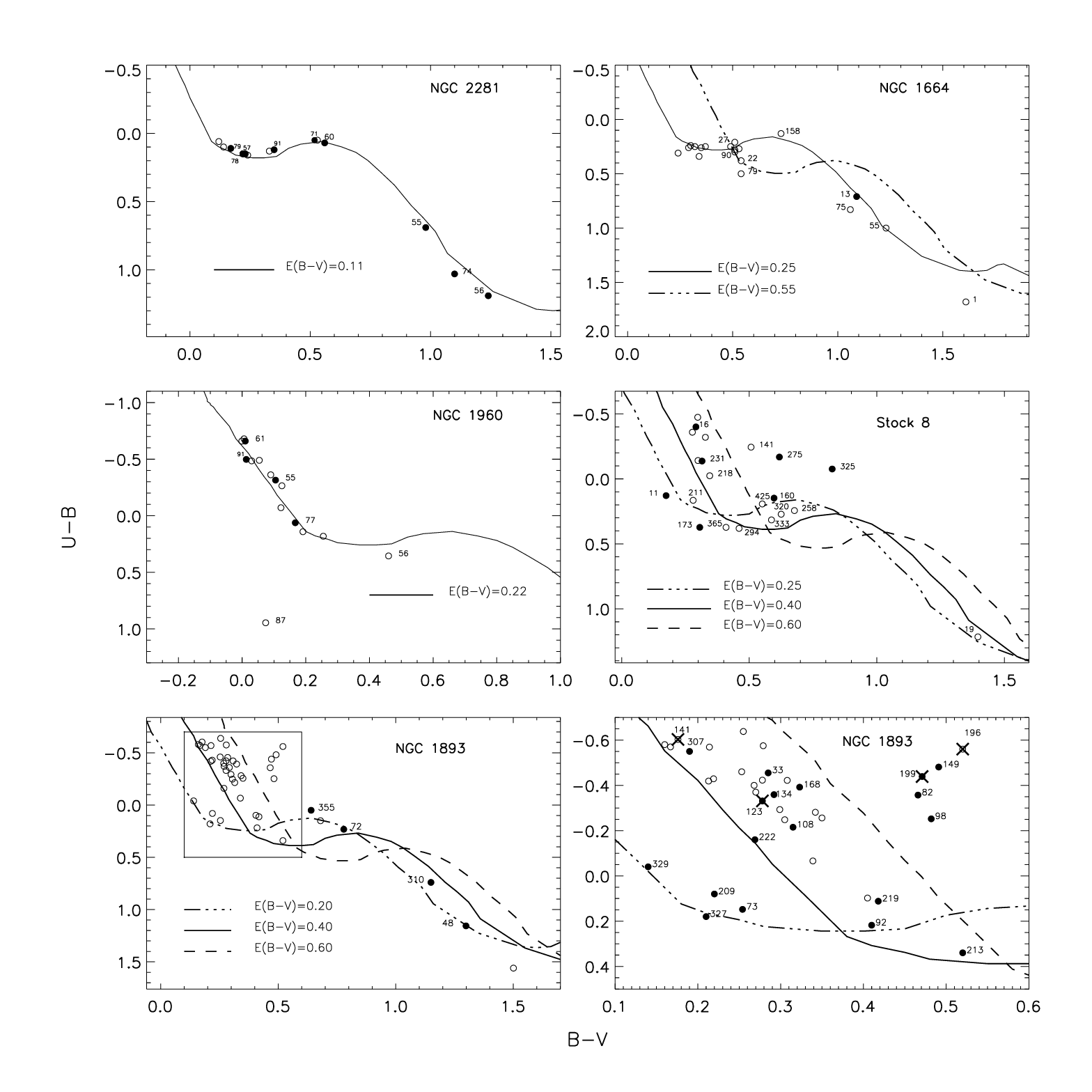}}
\caption{$(U-B)$ versus $(B-V)$ colour-colour diagram for the 
stars in the cluster region. Theoretical zero-age-main sequence (ZAMS) 
is taken from Schmidt-Kaler (1982) and is shifted along a reddening 
vector with an adopted slope of $E(U-B)/E(B-V)$=0.72, 
to match the observed colours. The symbols are same as in Fig. \ref{U_Q4}. 
In the case of NGC 1893, the H$\alpha$  emission stars are indicated with 
cross symbols. The lowermost-right  panel shows the enlarged view of the colour-colour diagram 
shown within the inset of NGC 1893 colour-colour diagram (lower left panel).}
\label{UBBVCCD}
\end{figure*}

\subsection{Member identification}
The individual Stokes parameters of the polarization vector of the $V$-band, 
{\it$P_{V}$}, given by $Q_{V}=P_{V}~\cos(2\theta_{V})$ and $U_{V}=P_{V}~\sin(2\theta_{V})$ 
are estimated for all the observed stars towards five clusters and presented on 
a $U_{V}$ versus $Q_{V}$ plot in Fig. \ref{U_Q4}. 
In such a diagram, members of  a star cluster are expected to group together, 
while non-members are expected to show a 
scattered distribution. The reason is that the measured polarization of a star 
depends on the column density of aligned dust grains that lie in front of the star and hence the 
degree of polarization would be similar, lower or higher depending on whether 
the star is a cluster member, a foreground or a background to the cluster. 
Likewise, the polarization angles of cluster members 
would be similar but could be different for foreground or background non-member stars 
as light from them could have contributions from different or additional dust components. 
Therefore $U_V-Q_V$ plot could be a useful tool to segregate the members and non-members
of a cluster. Stars with intrinsic polarization, e.g., 
due to an asymmetric distribution of matter around young stellar objects (YSOs) 
and/or rotation in their polarization angles 
(see section \ref{serkowski}) may also create scattered distribution 
in the $U_V-Q_V$ plane. 

As shown in Fig. \ref{U_Q4}, in the case of  nearby clusters $\it{viz}$ 
NGC 2281, NGC 1664 and NGC 1960, a grouping 
(bound visually by the dotted circle) is apparent, 
whereas in the case of distant clusters i.e. Stock 8 
and NGC 1893, the $U_{V}-Q_{V}$ diagram shows a scattered distribution. 
A scattered distribution in the case of a distant cluster is expected as the 
cluster region is contaminated by foreground/ background field stars. 
Moreover, clusters Stock 8 and NGC 1893 are young and have differential reddening, 
hence cluster matter may also affect the polarization. 
The apparent clustering in the case of NGC 2281, NGC 1664 and NGC 1960 further supports 
the above notion as these are relatively nearby clusters and have less field star contamination. 
Since these clusters are of intermediate age ($\ga$ 25 Myr), 
the effect of variable reddening is absent. 
The distribution is significantly scattered in 
the case of NGC 1893, which is the most distant cluster 
in the present study.  It seems difficult to identify probable 
cluster members of NGC 1893  using the $U_{V}-Q_{V}$ diagram as shown in Fig. \ref{U_Q4}. 
To further elucidate the membership in the clusters which do not show any apparent grouping 
we plot a box with dashed line in $U_{V}-Q_{V}$ plots (Fig. \ref{U_Q4}) 
having boundaries of mean $P_{V} \pm \sigma$ and mean $\theta_{V} \pm \sigma$ 
(as mentioned in section \ref{results}). 
It can be seen that the majority of stars of apparent grouping lie within 
the boundaries of the box as mentioned above. Therefore, in the case of 
clusters, where grouping is absent (Stock 8 and NGC 1893), we presume that 
the stars lying within the $1\sigma$ box of mean $P_{V}$ and $\theta_{V}$ values 
are probable members of the cluster. 

The combination of $U_{V}-Q_{V}$ diagram and $(U-B)/(B-V)$ 
colour-colour diagram can yield a better identification of probable 
members in a distant cluster (see e.g. Haikala 1995, Feinstein et al. 2008). 
Fig. \ref{UBBVCCD} shows the $(U-B)/(B-V)$ colour-colour diagram for 
NGC 2281, NGC 1664, NGC 1960, Stock 8 and NGC 1893 cluster regions. 
For the clusters NGC 2281, NGC 1664, NGC 1960 and Stock 8, 
the $(B-V)$ and $(U-B)$ colours are taken from the Tables \ref{result_2281}, \ref{result_1664}, 
\ref{result_1960} and \ref{result_stock8} respectively, whereas for the cluster NGC 1893, 
the colours are taken from Sharma et al. (2007) and Massey et al. (1995). 
The continuous curve represents the reddened 
ZAMS for the cluster region as per $E(B-V)$ 
values mentioned in Table \ref{basic_para}. In Fig. \ref{U_Q4}, 
the stars lying within the $1\sigma$ box are shown by open circles, whereas 
stars outside the $1\sigma$ box are shown by filled circles. It is apparent from 
Fig. \ref{U_Q4} that stars located within the $1\sigma$ box 
follow the general reddening of the cluster region, hence may be 
probable members of the clusters. 
On the basis of Figs \ref{U_Q4} and \ref{UBBVCCD}, we can draw the following conclusions:

NGC 2281: All the stars lying within the $1\sigma$ box and the boundary of apparent grouping 
follow the ZAMS reddened for $E(B-V)$ (=0.11 mag) of the cluster region. 
Although stars \#55, 56 and 74 
lie on the reddened 
ZAMS, they have $(B-V)$ $\ga$ 1.0. 
These stars could not be MS members of the cluster having log(age) = 8.7 yr. 
The star \#60 is located away from the grouping (Fig. \ref{U_Q4}). 
However it follows the reddened ZAMS and its colours are comparable to the probable MS members. 

NGC 1664: Stars located within the region bound by a dotted circle 
(see Fig. \ref{U_Q4}) have $P_{V}$ $\la $ 2\% and nicely follow 
the ZAMS reddened by  $E(B-V)$ = 0.25 mag. Hence these 
could be probable members of the cluster. 
The stars \#1, 22, 27 and 79 are located outside the circle but within the 
$1\sigma$ box. 
These stars show polarization in the range $2.0 \la P_{V} \la 2.5 \%$ and 
follow the ZAMS reddened by $E(B-V)$ $\sim$ 0.55 mag. 
Stars \#9, 13, 47 and 87 lie outside the $1\sigma$ box. 
These stars show 
$P_{V}$ values ranging from 2.5 $-$ 4.1 $\%$. 
These two groups of stars should be background stars. 
Star \#13 has $P_V$ = 2.59\%, 
but its estimated $E(B-V)$ $\sim$ 0.25 is comparable to the cluster's $E(B-V)$ value. 
As the age of the cluster is $\sim$ 500 Myr, the star \#13 ($V$=14.36, {\it{B-V}}=1.09) 
could not be a MS member. Star \#164, located significantly away from the grouping, 
is the lowest polarized star (0.97$\%$) and its polarization angle is 
significantly different (26$\pm$7$\degr$) from those of the remaining stars of the region. 
It should be a foreground non-member. 

NGC 1960: The stars located within the bound region and the 
$1\sigma$ box indicate $E(B-V)$ $\sim$ 0.22 mag. 
Star \#55 is located significantly away from the general distribution, 
but follows the ZAMS reddened by the average reddening of the region
(cf. Fig. \ref{UBBVCCD}). 
This star has relatively a high polarization ($P_{V}=1.44\pm0.24\%$) 
and small polarization angle ($\theta_{V}=147 \pm 5\degr$) in comparison to the 
remaining stars of the region.
It should be a background field star. 
Star \#61, the least polarized star ($P_{V}=0.81\%$) of the region, 
lies on the reddened ZAMS, but is located 
away from the general distribution. 
We conclude that star \#61 is probably a foreground field star. 
Although star \#77 is located beyond the 1$\sigma$ box, it follows the reddened 
ZAMS. This star has polarization $P_{V}=1.50\pm0.31\%$ and could be a background 
non-member. 
Star \#91 seems to be associated with the apparent grouping. Its polarization 
($P_{V}=1.25\pm0.14\%$) is comparable to the cluster mean $P_{V}$ value, hence 
it could be a member of the cluster. 

Stock 8: The $U_{V}-Q_{V}$ diagram does not show any grouping. 
The $(U-B)/(B-V)$ colour-colour diagram indicates that barring 
stars \#211 and 365, the remaining stars lying within the 1$\sigma$ box 
have $E(B-V)$ in the range of $0.40-0.60$ mag, hence should be members of the cluster. 
Stars \#11, 160 and 173 lie outside the 1$\sigma$ box. These three 
stars, along with stars \#19, 211 and 365, have $E(B-V)$ $\sim$ 0.25 mag. 
Hence these should be foreground stars. 
Stars \#16 and 231 are distributed near 
the edge of the $1\sigma$ box and have $E(B-V)$ and polarization values 
comparable to the mean values for the cluster region. These two stars could be 
cluster members. The stars \#275 and 325 are located out side the $1\sigma$ box, 
however they seem to be reddened B type stars of the region (cf. Fig. \ref{UBBVCCD}).
Star \#275 has significantly different polarization angle 
(136$\pm$5$\degr$) from the mean value for the cluster region ($160\pm7$$\degr$), 
although the polarization value ($2.37\pm0.39\%$) is similar to the cluster 
mean value (2.40$\pm$0.60$\%$). 
Star \#325 has a relatively higher $P_V$ (4.18$\pm$0.44$\%$) value than 
the mean value for the cluster region (2.40$\pm$0.60$\%$) although the polarization angle 
(165$\pm$3$\degr$) is similar to the cluster mean value (160$\pm$7$\degr$). 
Hence the membership determination of these two stars, \#275 and 325, is uncertain. 
The approximate boundary 
to demarcate the foreground stars from the cluster 
members is shown by a thin continuous curve (see Fig. \ref{U_Q4}).\\

NGC 1893: The colour-colour diagram of the region reveals that 
barring stars \#197 and 356, all the stars lying within the $1\sigma$ 
box have $E(B-V)$ $\sim$ 0.4$-$0.6 mag, indicating that these 
stars may be cluster members. Fig. \ref{UBBVCCD} reveals that stars 
\#48, 72, 73, 92, 209, 310, 327, 329 and 355 have $E(B-V)$ 
$\sim$ 0.2 mag, hence these should be field stars. 
Five stars \#33, 134, 168, 219 and 222 are located 
near the boundary of the $1\sigma$ box. 
These stars have $E(B-V)$ in the range of 0.40$-$0.60 mag, hence 
these could also be members of the cluster. Star \#149 is apparently located 
away from the distribution. This star has relatively higher value of 
$\overline\epsilon$ (2.43; cf. Sec. 4.2 and Table \ref{tab_dust_prop}) and may have 
a rotation in the polarization angle. Two stars \#141 (S2R2N43, Marco and Negueruela 2002) 
and \#196 (S1R2N35, Negueruela et al. 2007)
are $H\alpha$ emission stars. On the basis of polarization, reddening and 
position in the $U_V-Q_V$ plot, star \#141 should be a cluster member. 
However, star \#196 is a highly reddened MS star ($E(B-V)$=0.83) and 
is probably an HAe/Be star (see Fig. \ref{NIRCCD}), 
but its $P_{V}$ ($2.45\pm0.29\%$) and $\theta_{V}$ ($160\pm3\degr$) 
suggest that it could have undergone depolarization effect. Hence, the membership of star 
\#196 is uncertain. 
The stars having relatively high polarization and high $E(B-V)$ values 
as well as having intrinsic component of polarization have been demarcated 
using a dotted curve as shown 
in Fig. \ref{U_Q4}. Fig. \ref{UBBVCCD} indicates that stars \#82 and 98 
are highly reddened stars, 
whereas 123 and 199 are $H\alpha$ 
emission stars with possible intrinsic component of polarization. 
Star \#108 seems to be a classical Be star (cf. Fig. \ref{NIRCCD}) and may have 
an intrinsic component of polarization. 
Star \#213 may have rotation in its polarization angle as its
$\overline \epsilon=1.45$ (see Table \ref{tab_dust_prop}). 

The probable members of the cluster identified using $U_{V}-Q_{V}$ and colour-colour 
diagrams, along with the kinematic membership probability ($M_{P}$) taken from Vasileviskis \& Balz (1959) 
for NGC 2281 and from Dias et al. (2006) for remaining clusters, 
are mentioned in Tables \ref{result_2281}, \ref{result_1664}, 
\ref{result_1960}, \ref{result_stock8} and \ref{tab_dust_prop}. 
Stars with kinematic membership probability $\geq$ 50$\%$ are considered as members.
The member and non-member stars are represented with {\it{M}} and {\it{NM}} respectively. 
Stars with uncertainty in their membership determination are indicated with a '?' symbol. 
A comparison indicates that the discrepancy between the membership 
estimated from kinematic criterion and in the present work is $\sim 15\% - 20\%$ 
for the nearby clusters (i.e., NGC 2281, NGC 1664 and NGC 1960). 
The discrepancy is found to increase 
($\sim 30\% - 45\%$) for distant clusters (i.e., Stock 8 and NGC 1893). 
The mean values of  $P_V$ and $\theta_V$ of member stars of the respective 
clusters as identified above as well as of field stars lying in the cluster 
regions are given in Table \ref{Mean_M_NM}. The  mean  values of $P_V$ 
for members of the two nearby clusters $\it viz$  NGC 2281 and NGC 1960 
are the same as those for 
field stars in these regions, whereas in the case of other nearby cluster 
(NGC 1664), the non-members show higher value for mean $P_{V}$.  
The higher values of $P_V$ for field stars towards the direction 
of cluster NGC 1664 are due to the fact that the non-member field 
stars are located in the background of the cluster and have higher extinction 
(see Fig. \ref{UBBVCCD}). In the case of the distant clusters Stock 8 and NGC 
1893, the mean values of $P_{V}$ for non-members are lower than those for member stars. 
The mean values of $P_{V}$ obtained for cluster members further manifest 
an increasing trend with distance. Table \ref{Mean_M_NM} also indicates that 
the mean $\theta_{V}$ values are almost the same for members and non-members 
lying towards the direction of the clusters studied in the present work. 
\begin{figure*}
\resizebox{12cm}{12cm}{\includegraphics{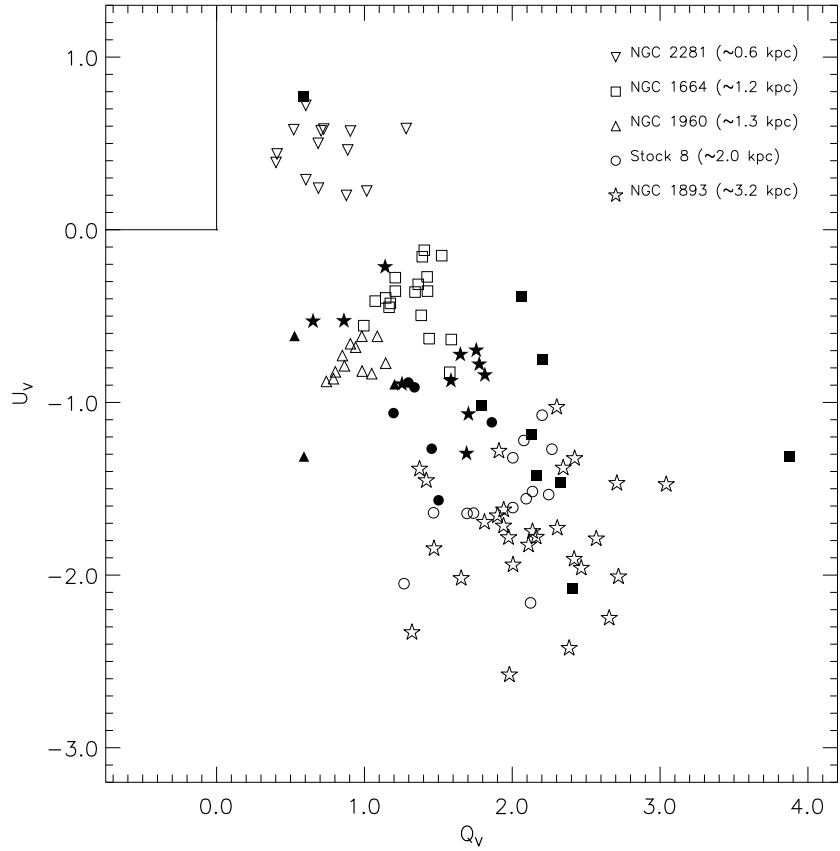}}
\caption{The $U_{V}$ versus $Q_{V}$ diagram for stars lying in the cluster regions studied,
namely, NGC 2281 (inverted triangles), NGC 1664 (squares), NGC 1960 (triangles), Stock 8 (circles) and
NGC 1893 (star symbols). The filled symbols represent field stars in the respective cluster regions.
The rectangular box covered by $Q_{V}=0$ and $U_{V}=0$ is the dust free environment of 
the solar neighbourhood.}
\label{U_Q}
\end{figure*}

The $U_V-Q_V$ plot is also a useful tool to study the interstellar dust
distribution as a function of distance from the Sun to the clusters (e.g., Feinstein et al. 2008). 
The clusters studied in the present work have distances 
in the range of $\sim 0.6$ to $\sim3$ kpc. 
The polarimetric results obtained can be 
used to study 
the properties of the dust
towards the anticenter direction ($l$ $\sim160{\degr}$ - $175\degr$) of the Galaxy. 
Based upon the earlier discussion using  $Q_V-U_V$ and $(U-B)/(B-V)$ 
diagrams, in Fig. \ref{U_Q} we show a combined $U_V-Q_V$ diagram for all the five clusters, namely, 
NGC 2281 (inverted triangles), NGC 1664 (squares), NGC 1960 (triangles), Stock 8 (circles) 
and NGC 1893 (star symbols). 
Open and filled symbols are used to represent members and non-members based upon our analysis. 
The dust free environment of the solar neighbourhood (shown with rectangular box) 
is represented by $Q_{V}=0$ and $U_{V}=0$, while any other point on this plot 
represents the direction of the polarization 
vector seen towards that direction from the Sun. 
Fig. \ref{U_Q} suggests that the degree of polarization of stars is found to increase 
with the distance to the clusters, as also noticed in Fig. \ref{Hist_5}. 
This is consistent with the fact that the degree of polarization increases 
with the column density of dust grains lying in front of the stars that are relatively well aligned. 
The close proximity of the points of NGC 1664 and NGC $1960$ in $U_V-Q_V$ plane is consistent with 
the fact that they are located approximately at similar distances (1.2 and 1.3 kpc, respectively).

\subsection{The Serkowski law}\label{serkowski}
The wavelength dependence of polarization towards many Galactic directions 
follows the Serkowski's law of interstellar polarization 
(Serkowski 1973; Coyne, Gehrels \& Serkowski 1974; Wilking, Lebofsky, \& Rieke 1982): 
\begin{equation}\label{serkowski_law}
 P_{\lambda} = P_{max} \exp[-K~\ln^{2} (\lambda_{max}/\lambda)]
\end{equation}
Where $P_{\lambda}$ is the percentage polarization at wavelength $\lambda$
and $P_{max}$ is the peak polarization, occurring at wavelength $\lambda_{max}$.
The $\lambda_{max}$  is a function of optical properties
and characteristic particle size distribution of aligned grains
(Serkowski, Mathewson \& Ford, 1975; McMillan, 1978).
The value of $P_{max}$ is determined by the column density,
the chemical composition,
size, shape, and 
alignment efficiency of the dust grains.
The parameter $K$, an inverse measure of the width of the polarization curve, 
was treated as a constant by Serkowski et al. (1975), who adopted a value of 1.15
for all the stars. The Serkowski's relation with $K$=1.15 provides an adequate
representation of the observations of interstellar polarization between
wavelengths 0.36 and 1.0 $\mu$m. If the polarization is 
produced by aligned interstellar dust grains, the observed data will follow equation 
(\ref{serkowski_law}) and hence we can estimate $P_{max}$ and $\lambda_{max}$ for each star. 
The $P_{max}$ and $\lambda_{max}$ are obtained using the weighted least-squares fitting 
to the measured polarization in $BV(RI)_{c}$ bands to equation (\ref{serkowski_law}) 
by adopting $K$=1.15. We have also computed the parameter 
$\sigma_{1}$\footnote{The values of $\sigma_{1}$ 
for each star are computed using the expression 
$\sigma_{1}^{2}=\sum(r_{\lambda}/\epsilon_{p\lambda})^{2}/(m-2)$; 
where $m$ is the number of colours and 
$r_{\lambda}=P_{\lambda}-P_{max} \exp[-K~ln^{2} (\lambda_{max}/\lambda)$.} 
(the unit weight error of the fit) for each star that quantifies the departure 
of the data points from the standard Serkowski's law. Because of the weighting scheme, 
the values of $\sigma_{1}$ should not exceed 1.5, but if they do, it implies that the stars 
have intrinsic polarization (e.g., Feinstein et al. 2008, Medhi et al 2007, 2008, 2010). 
The $\lambda_{max}$ values can also be used to infer the origin of the polarization. 
The stars with $\lambda_{max}$ much lower than the average value of the 
interstellar medium (0.545 $\mu$m; Serkowski, Mathewson \& Ford 1975) 
may have an intrinsic component of polarization. 
Another parameter to infer the presence 
of intrinsic polarization or polarization angle rotation along the line of sight 
(Coyne 1974, Martin 1974), is the dispersion of the polarization angle for each 
star normalized by the average of the polarization angle errors, 
$\overline{\epsilon}$ (Marraco et al. 1993; Orsatti et al. 2007; Feinstein et al. 2008).

The estimated values of $P_{max}$, $\lambda_{max}$, $\sigma_{1}$ and $\overline{\epsilon}$ 
for the stars towards NGC 1893 are given in Table \ref{tab_dust_prop}. 
The star identifications are same as given in Table \ref{result_1893}. 
The weighted mean values of the $P_{max}$ and $\lambda_{max}$ are found to 
be $2.59 \pm 0.02\%$ and  $0.55 \pm 0.01$ $\mu$m respectively. 
The estimated $\lambda_{max}$ is quite similar to the value corresponding 
to the general interstellar medium (0.545 $\mu$m, Serkowski et al. 1975). 
Using the relation $R_{V}=5.6\times\lambda_{max}$ 
(Whittet \& Van Breda 1978), the value of $R_{V}$, 
the total-to-selective extinction, comes out to be 
3.08$\pm$0.05, which is in agreement with the average 
value ($R_{V}$ = 3.1) for the Milky Way Galaxy, indicating 
that the size of the dust grains 
within the cluster NGC 1893 is normal. 
Similar conclusion was drawn by Sharma et al. (2007) using 
$(\lambda-V)/(B-V)$ two colour diagrams. 
\begin{figure*}
\resizebox{12cm}{18cm}{\includegraphics{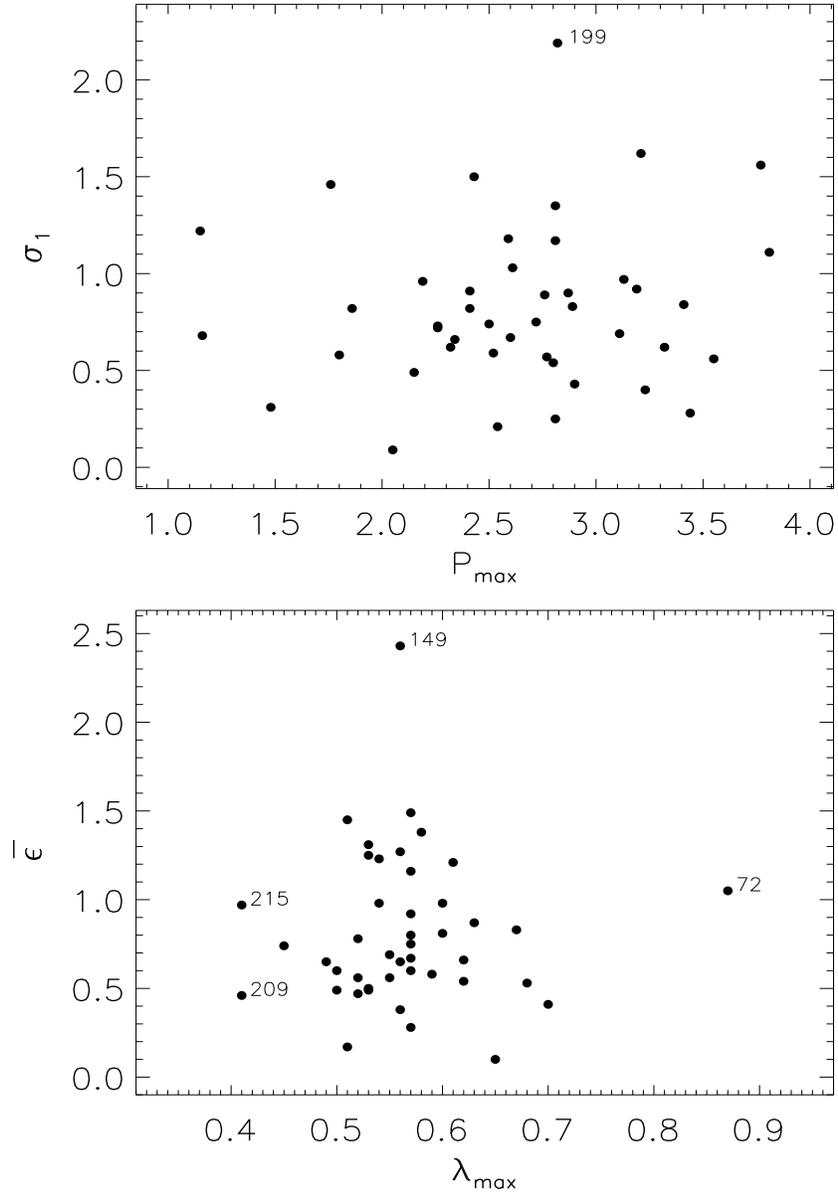}}
\caption{Top panel is the plot between $\sigma_{1}$ and $P_{max}$, whereas the bottom panel
is the plot between $\overline{\epsilon}$ versus $\lambda_{max}$.}\label{Pmax_Lmax} 
\end{figure*}
In Fig. \ref{Pmax_Lmax} we show $\sigma_{1}$ versus P$_{max}$ (upper panel) 
and $\overline{\epsilon}$ versus $\lambda_{max}$ (lower panel) plots. 
The criteria mentioned above indicate that majority of the 
stars do not show evidence of intrinsic polarization. 
However a few stars \#72, 149, 199, 209 and 215 
show deviation from the general distribution. 

\begin{figure*}
\resizebox{16cm}{13cm}{\includegraphics{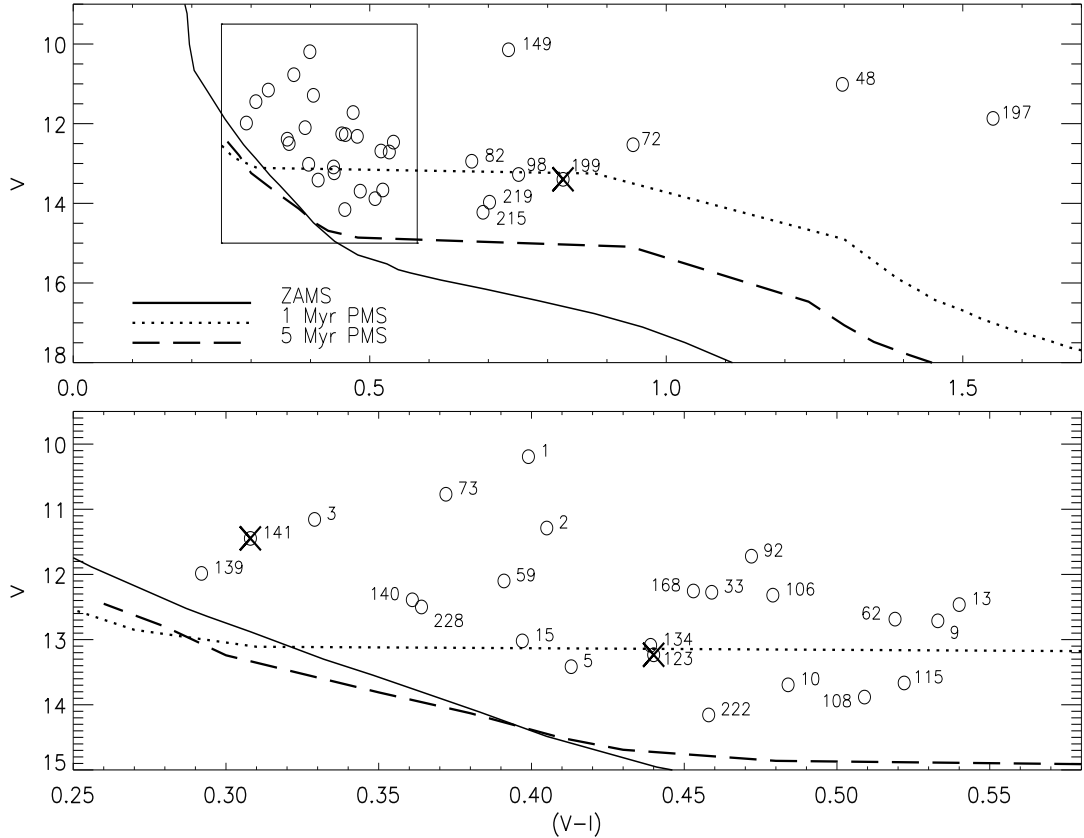}}
\caption{Upper panel is the $V$ versus $(V-I)$ colour-magnitude diagram 
for the NGC 1893 region. The continuous, dotted and dashed curves represent the ZAMS from 
Girardi et al. (2002) and PMS isochrones for 1 and 5 Myr by Siess et al. (2000) respectively. 
The isochrones are adjusted for a distance of 3.2 kpc and $E(B-V)$=0.40 mag. 
The stars having $H\alpha$ emission are 
shown by crosses. The lower panel is the enlarged view of the colour-magnitude diagram 
shown in the inset of the upper panel.} \label{VIVCMD}
\end{figure*}

\begin{figure*}
\resizebox{12cm}{13cm}{\includegraphics{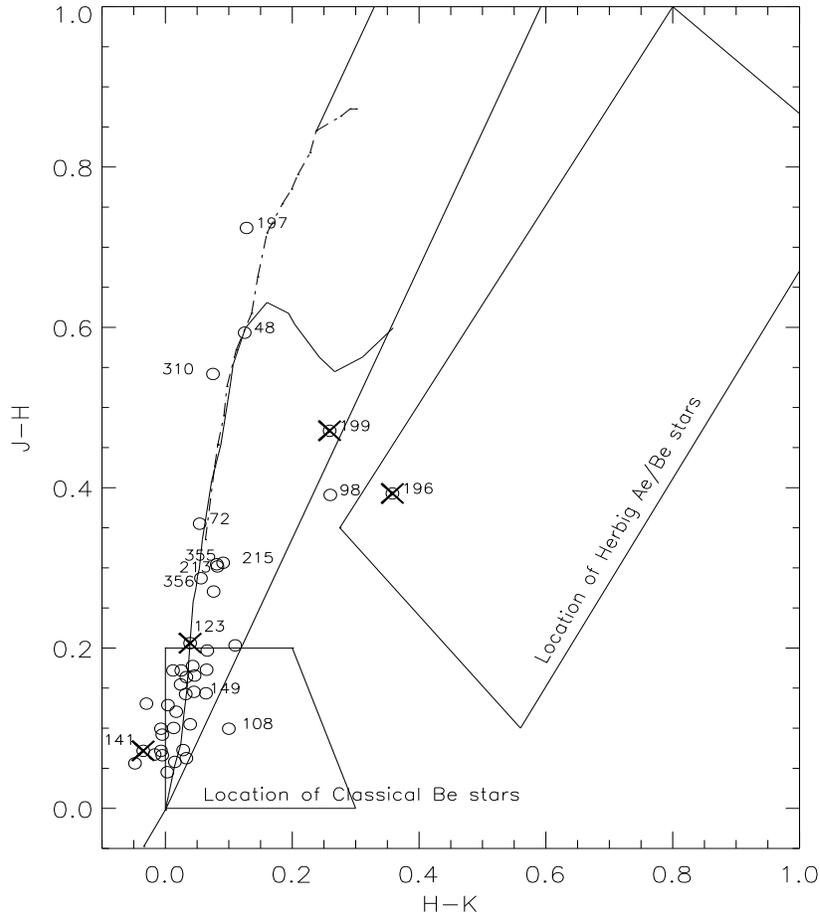}}
\caption{$(J-H)$ versus $(H-K)$ colour-colour diagram for the NGC 1893 region. 
The data is taken from the Two-Micron All-Sky Survey (2MASS) Point Source Catalog (PSC) 
(Cutri et al. 2003). 
2MASS data has been converted in to California Institute of Technology (CIT) system using 
the relations provided by Carpenter (2001). The theoretical tracks for dwarfs and giants 
are drawn (Bessell \& Brett, 1988). 
Reddening vectors are also drawn (Cohen et al. 1981). 
The location of Be stars (cf. Dougherty et al. 1994), 
and the location of Herbig Ae/Be stars (cf. Hernandez et al. 2005) are also shown. 
The stars having $H\alpha$ emission 
are shown by crosses.}\label{NIRCCD}
\end{figure*}

Star \#199 is an $H\alpha$ 
emission star and its location 
in the $V/(V-I)$ colour-magnitude diagram (CMD) (Fig. \ref{VIVCMD}) reveals that 
it should be a pre-main sequence (PMS) star. The location of star \#209 in 
the $(U-B)/(B-V)$ colour-colour diagram 
(Fig. \ref{UBBVCCD}) suggests that it could be a foreground field star. 
The $V/(V-I)$ CMD reveals that star \#215 should be a PMS star. 
Star \#123 and 199 are identified as $H{\alpha}$ emission stars 
S1R2N44 and S1R2N56, respectively (Neguerela et al. 2007).
Stars \#199 and 215 are likely to be PMS stars and accreting matter from their 
circumstellar material. The accreted material, 
probably distributed in an asymmetric disk geometry, 
might be responsible for the observed 
intrinsic component of polarization.
The computed values of $\overline{\epsilon}$ 
show no significant rotation (except for star \#149) 
in polarization angles reiterating the fact that the dust component responsible 
for the measured polarization is well aligned. 
Star \#149 is a highly reddened 
MS star (see. Figs \ref{UBBVCCD}, \ref{VIVCMD} and \ref{NIRCCD})
as it is embedded at the eastern edge of the wind
blown bubble (cf. Sharma et al. 2007).
\begin{figure*}
\resizebox{15cm}{13cm}{\includegraphics{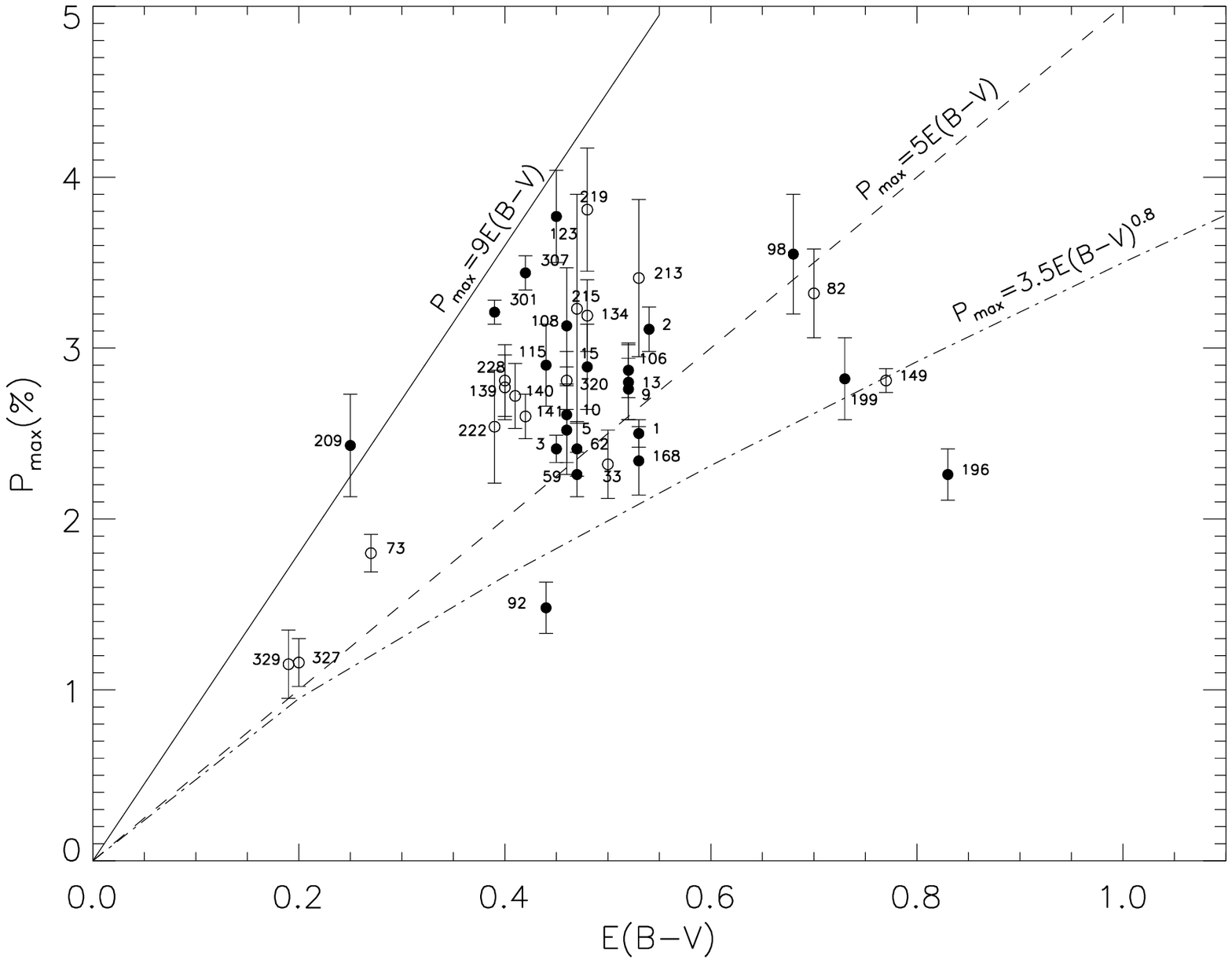}}
\caption{ The polarization efficiency diagram. 
Filled and open circles are the stars distributed 
in the northern and southern regions of the cluster respectively. 
The solid line represents the empirical
upper limit relation for the polarization efficiency of 
$P_{max}=9\times E(B-V)$ (Serkowski et al. 1975). 
The dashed line represents the relation $P_{max}=5\times E(B-V)$ (Serkowski et al. 1975) 
and the dashed-dotted line represents the relation 
$P_{max}=3.5\times E(B-V)^{0.8}$ by Fosalba et al. (2002).}\label{pmaxebvpoleffi}
\end{figure*}

\subsection{Polarization efficiency}\label{pol_effi}
The degree of polarization produced for a given amount of 
extinction (or reddening) is referred as the 
polarization efficiency of the intervening dust grains. 
The polarization efficiency depends mainly on 
the orientation of the magnetic field along the line of sight, 
the magnetic field strength and the degree of alignment of the dust grains.
Mie calculations place a theoretical upper limit of $P_{max}=43\times E(B-V)$ 
(Whittet 1992 and references there in) on the polarization efficiency by 
an infinite cylinder with the diameter comparable to the 
wavelength of the incident light and with their long axes parallel 
to each other and perpendicular to the line of sight. 
The empirical upper limit relation for the polarization efficiency 
resulting from the studies of reddened Galactic stars, 
assuming normal interstellar material
characterized by $R_{V}=3.1$ (Serkowski, Mathewson \& Ford 1975) 
is found to be $P_{max}=9\times E(B-V)$. 
The fact that the maximum observed polarization efficiency is found to be 
less by a factor of 4.8 than that expected from theory, 
implies that the alignment of dust grains is not 
perfect. The reason could be the 
presence of various components of magnetic fields 
oriented differently along the line of sight and/or the grains 
are only moderately elongated rather than infinite cylinder.

The $UBV$ photometric data by Sharma et al. (2007) and Massey et al. (1995) 
have been used to estimate the reddening $E(B-V)$. The reddening 
of individual stars having spectral type earlier than A0, has been 
derived using the $Q$ method (Johnson \& Morgan 1953) and the values 
are given in Table \ref{tab_dust_prop}. 
The $(J-H)/(H-K)$ near infrared (NIR) colour-colour diagram (Fig. \ref{NIRCCD}) 
reveals that stars \#48, 72, 197, 310, 355 and 356 have 
spectral type later than A0. The $(U-B)/(B-V)$ colour-colour 
diagram (Fig. \ref{UBBVCCD}) indicates that these stars follow 
$ZAMS$ reddened by $E(B-V)$=0.2 mag, hence we assign 
reddening of $E(B-V)$=0.20 mag. 

In Fig. \ref{pmaxebvpoleffi}, we present $P_{max}$ versus $E(B-V)$ 
for stars towards NGC 1893 region. The continuous line shows the empirical 
upper limit for the polarization efficiency given 
by $P_{max}=9\times E(B-V)$ (Serkowski et al. 1975), 
assuming normal interstellar material characterized by $R_{V}=3.1$. 
The recent estimate of the average efficiency 
by Fosalba et al. (2002), which
is valid for $E(B-V) \textless$ 1.0 mag, is represented by the 
relation $P_{max}=3.5\times E(B-V)^{0.8}$ 
and is shown by the dash-dotted line.
For comparison, the average polarization efficiency relation, 
$P_{max}=5\times E(B-V)$ (Serkowski et al. 1975), 
is also drawn using the dashed line. 
A majority of the points are found to lie between the continuous line 
and above the dashed line, indicating that the dust grains towards NGC 1893 
have higher polarization efficiency in comparison to the the general ISM. 

\begin{figure*}
\resizebox{14cm}{16cm}{\includegraphics{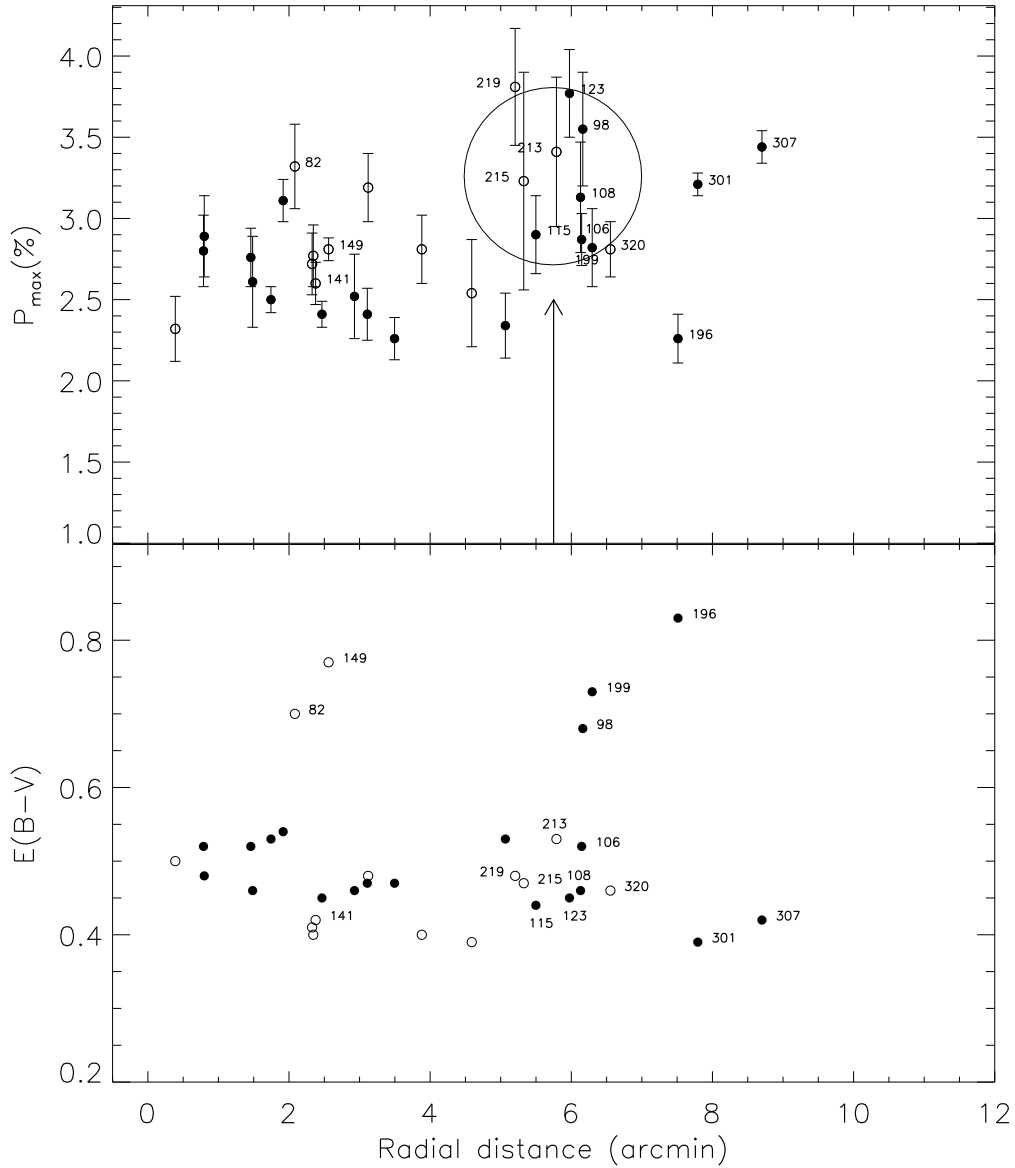}}
\caption{The radial variation of $P_{max}$ (upper panel) and $E(B-V)$ (lower panel) 
for the cluster members and the stars (\# 108, 123, 196, 199 and 213)
with uncertainty in their membership. 
Filled circles are the stars distributed in the northern part, whereas open circles are 
in the southern part of the cluster NGC 1893. 
Most of the stars falling within the circle at a distance of $5\farcm5$ are 
distributed nearer the two nebulae Sim 129 and 130.
}
\label{Pmax_ebv_raddist}
\end{figure*}
\begin{figure*}
\resizebox{10cm}{10cm}{\includegraphics{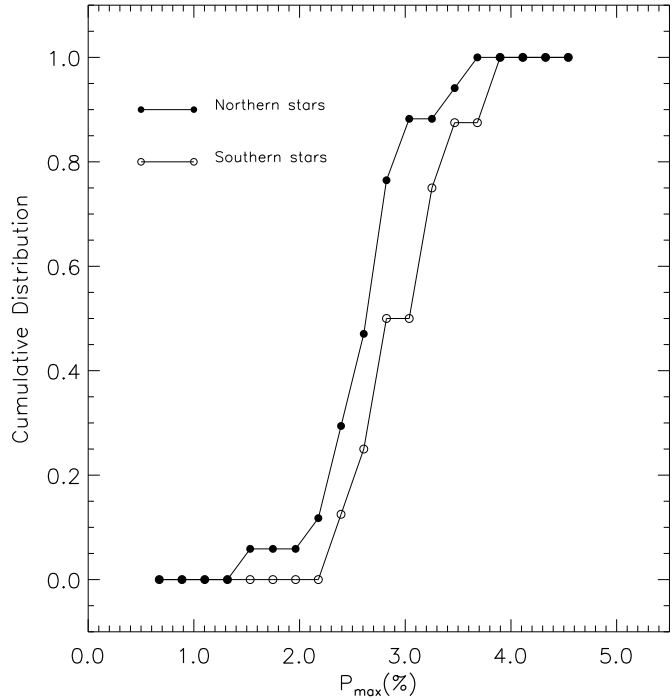}}
\caption{Cumulative distribution of $P_{max}$ for the stars, 
with $E(B-V)$ values between 0.4-0.6 mag, distributed in 
northern and southern parts of the cluster NGC 1893. 
The lines connecting filled and open circles are 
the cumulative distributions for northern and southern stars respectively.}\label{Cumul_distri_Pmax_lmax}
\end{figure*}

\subsection{Spatial variation of $P_{max}$ and $E(B-V)$}\label{Spatial_vari_pmax_ebv}

The upper panel of Fig. \ref{Pmax_ebv_raddist} shows $P_{max}$ 
for the cluster members and the stars 
with uncertainty in their membership (\# 108, 123, 196, 199 and 213), 
as a function of the radial distance 
from the ionization source (see Fig. \ref{NGC1893V_band}: HD 242935; shown by the plus symbol). 
Stars located in the northern and southern 
parts of the cluster are shown by filled and open circles respectively. 
The distribution of $P_{max}$ 
within a radial distance of $\sim$ 5$^{\arcmin}$ from the ionization source shows 
a decreasing trend with the increase in radial distance. The distribution of $P_{max}$ 
shown in Fig. \ref{Pmax_ebv_raddist} also reveal that $P_{max}$ in the 
southern region is relatively higher than that of 
the northern region. 
There is a significant increase 
in the $P_{max}$ values at $\sim$ $5\farcm5$. 
The increase in $P_{max}$ is mainly due to the YSOs distributed around 
the nebulae Sim 129 and 130, 
located towards the North-East direction of 
HD 242935 at a radial distance of $\sim 6^{\arcmin}$.

The lower panel of Fig. \ref{Pmax_ebv_raddist} shows radial variation of 
$E(B-V)$. The distribution 
of $E(B-V)$ reveals a clumpy nature of the gas/dust distribution. 
The polarization efficiency in the northern region of the cluster at a 
radial distance $\sim$ $0\farcm5$, 
assuming a normal value of $R_{V}$=3.1, comes out to be $\sim 6.0$, 
and decreases to $\sim 5.0$ at the boundary ($\sim 4^{\arcmin}$) of the cluster. 
The ratio $P_{max}/ E(B-V)$ depends mainly on 
the alignment efficiency, magnetic strength and amount of depolarization 
due to radiation traversing through more than one medium. The polarization efficiency 
$P_\lambda/A_\lambda$ also depends on the particle shape (Voshchinnikov \& Das 2008). 

The weighted mean of $P_{max}$ for stars having 0.40 $\leq E(B-V) \leq 0.60$ lying in 
the northern and southern regions of the cluster is estimated to be as 
2.64$\pm$0.04$\%$ and 2.77$\pm$0.07$\%$. 
Fig. \ref{Cumul_distri_Pmax_lmax} shows 
the cumulative distribution of $P_{max}$ in both regions, for
stars having 0.40 $\leq E(B-V) \leq 0.60$ mag, which indicates that
$P_{max}$ values in the southern region are systematically higher than
those in the northern region. 
Kolmogorov-Smirnov test indicates that these
distributions are different at 80\% confidence level.

The weighted mean values of polarization angle ($\theta_{V}$) 
for northern and southern regions are $161\pm4\degr$ and $162\pm6\degr$, 
which indicate that there is no difference in $\theta_{V}$ in the two regions.

\section{Dust components responsible for the observed polarization}\label{dust_components}

The degree of polarization of a star increases 
as a function of distance due to the presence of a column of aligned dust grains 
along the pencil beam of radiation from the star. 
The degree of polarization shows a sudden jump if the radiation from 
the stars encounter a dust layer located at a certain distance 
between the star and the observer. The number of such sudden jumps is 
characterized by the number of dust layers encountered by the radiation along 
its path and the relative magnetic field orientations in the dust layers.    

\begin{figure*}
\resizebox{11cm}{15cm}{\includegraphics{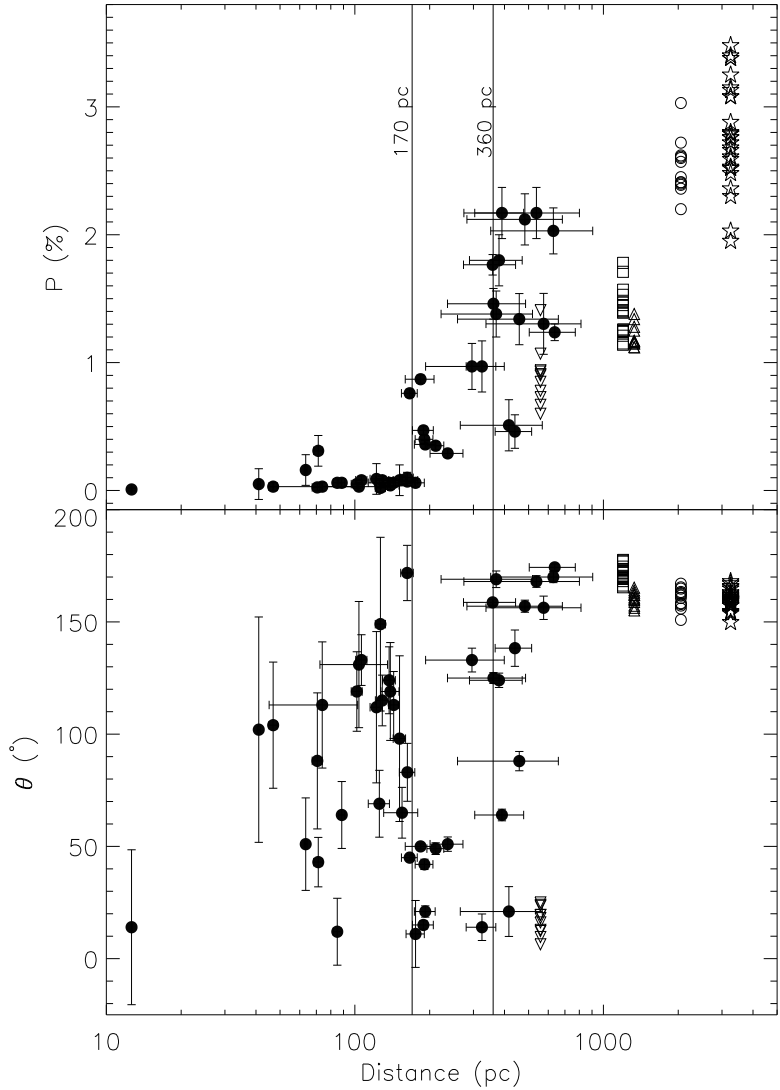}}
\caption{The change in $P_{V} (\%)$ and $\theta_{V} (\degr)$ as a function of distance (pc) 
is plotted in the upper and lower panels for stars (filled circles) selected 
from a circular region 
of $10\degr$ around NGC 1893. The $P\%$ and $\theta\degr$ values are obtained from the 
Heiles (2000) catalog and the distances are obtained from Hipparcos parallaxes (van Leeuwen, 2007). 
The vertical solid lines are drawn at 170 pc and 360 pc. The results from our study 
of the five clusters are also shown. The clusters are represented by the same symbols 
as in Fig. \ref{U_Q}.}\label{pol_PA_dist}
\end{figure*}

To understand the distribution of dust layers towards the direction of NGC 1893,
we selected stars from a region of 10${\degr}$ radius around the cluster
with both polarization measurements (Heiles 2000)
and Hipparcos parallaxes (van Leeuwen, 2007) available.
In Fig. \ref{pol_PA_dist} we show the degree of polarization versus distance (upper panel) 
and polarization angle versus distance (lower panel) plots for the selected stars (filled circles). 
Stars observed towards the clusters NGC 2281, NGC 1664, NGC 1960, Stock 8 and NGC 1893 
are also shown using inverted triangles, squares, triangles, open circles and star symbols 
respectively. Two significant jumps in the values of $P_{V}$, one at $\sim170$ pc and another 
at $\sim360$ pc, are clearly evident in the $P_{V}$ versus distance plot (upper panel). 

For distances $\la 170$ pc, the $\theta_{V}$ values show a scattered 
distribution between $\sim 50\degr - 130\degr$. At $\sim 170$ pc, the 
$\theta_{V}$ values show a dip to $\sim 20\degr$. A sharp rise in 
$\theta_{V}$ values from $\sim$20$\degr$ to 160$\degr$ is evident at a distance 
of $\sim 360$ pc. The stars located beyond $\sim 1000$ pc show an average 
polarization angle of $\theta_{V}$ $\sim$ $163\degr$. 
On the basis of Fig. \ref{pol_PA_dist}, we infer that 
a dust layer at $\sim 170 $ pc contributes $\sim 0.3-0.9\%$ to the polarization 
and the dust grains are aligned towards $\theta_{V}$ $\sim$ $20\degr$. Another 
dust layer with $\theta_{V}$  $\sim 160-170\degr$ at a distance $\sim 360$ pc 
further contributes $\sim 0.3-1.3\%$ (i.e., total $\sim$ 1.2 to $\sim$ 2.2 $\%$).
The clusters located at distance $\ga$ 1000 pc show $P_{V}$ 
in the range of $\sim 1-4\%$. The $P_{V}$ $\ga$ 2.2$\%$ in these clusters 
must be due to intracluster medium. The optical properties derived 
in Sections \ref{serkowski} and \ref{pol_effi} should be a combined 
effect of dust layers at d $\sim 170$ pc, $\sim 360$ pc and intracluster medium.

Neckel and Klare (1980) have studied $A_V$ distribution in the Galactic plane with
$|b| \leq 7.6\degr$ using extinction and distances computed
for individual stars. The $A_V$ map
towards the direction of NGC 1893 by Neckel and Klare (1980)
reveals that $A_V$ increases with distance
up to $\sim$ 2 kpc. Beyond $\sim$ 2 kpc $A_V$ is found to
be rather constant, indicating that the dust contribution in the distance range $2-3$ kpc
is negligible.
Here, it is worthwhile to mention that the mean polarization
$P_V$ in the Stock 8 (2.4$\pm0.6\%$) and NGC 1893 (2.6$\pm0.7\%$) is almost the same,
which further confirms that the contribution of dust in the distance
range $\sim$ 2-3 kpc is negligible.
Further, the observed mean polarization values are consistent with the
$E(B-V)$ values mentioned in Table \ref{basic_para}.

\section {Conclusions}\label{conclude}

We have made polarimetric observations of the open cluster NGC 
1893 in  $B$, $V$, $R_{c}$ and $I_{c}$ bands. 
Three additional open clusters towards the direction of NGC 1893, 
namely, NGC 2281, NGC 1960 and Stock 8 along with NGC 1664 
toward anticenter direction of the Galaxy were observed in $V$-band only.
The main aim of the study is to investigate the properties of dust grains 
towards the anticenter direction of the Galaxy between $l\sim160\degr$ to $\sim175\degr$, 
using stars of open clusters located in the distance 
range of 600 pc (NGC 2281) to 3.2 kpc (NGC 1893). 

The stars located at distances $\leq 170$ pc show a large scatter in 
$\theta_{V}$ values with a polarization of $\sim 0.1\%$. The degree 
of polarization is found to increase with the distance. The distribution 
of $P_{V}$ and $\theta_{V}$ as a function of distance reveals two dust 
layers at $\sim 170$ pc and $\sim 360$ pc. 
The first dust layer is characterized by its 
polarized components as $P_{V}$ $\sim 0.3-0.9\%$ and $\theta_{V}$ $\sim 20-50\degr$. 
Both dust layers produce a maximum combined polarization up to $P_{V}$ $\sim 2.2\%$. 
Polarization values higher than the $\sim 2.2 \%$ should be due to intracluster medium.

The magnetic field orientation remains unchanged within the Galactic plane $|b|<2\degr$. 
The estimated mean value of the polarization angle 
(assuming a Gaussian distribution) of 
the four clusters (NGC 1664, NGC 1960, Stock 8 and NGC 1893) comes out to be 
$\sim163\degr$ with a standard deviation of 6$\degr$. 
This small dispersion in polarization 
angle could be due to the presence of a uniform dust layer 
beyond 1 kpc. 
Present observations reveal that in case of NGC 1893,
the foreground two dust layers, in addition to the intracluster medium,
seems to be responsible for the polarization effects.

The estimated mean values of $P_{V}$ for the two clusters namely, Stock 8 and NGC 1893, imply 
that the ISM located between  2-3 kpc distance has negligible contribution 
towards extinction as well as the observed polarization. 
Present polarimetric results are consistent with the 
reddening distribution given by Neckel and Klare (1980). 

The weighted mean of the $P_{max}$ and $\lambda_{max}$ values for NGC 1893 are found 
to be $2.59 \pm 0.02\%$ and  $0.55 \pm 0.01$ $\mu$m respectively. 
The estimated $\lambda_{max}$ is quite similar to that of the general interstellar medium. 
The value of $R_{V}$ using the relation 
$R_{V}=5.6\times\lambda_{max}$
(Whittet \& Van Breda 1978) is found to be close to the average value of 3.1 
for the Milky Way Galaxy, implying that 
the average size of the dust grains within the cluster 
NGC 1893 is similar 
to the general interstellar medium. 
We also identified four candidate stars with intrinsic polarization in NGC 1893.

The radial distribution of $P_{max}$ within the cluster shows a decrease 
of $P_{max}$ towards the outer region of the cluster. The $P_{max}$ 
and the polarization efficiency is found to be higher towards the southern region of 
the cluster. 
We have shown that the polarization measurements in combination with 
$(U-B)/(B-V)$ colour-colour diagram provide a good tool to determine the membership 
in a cluster. 

\section*{Acknowledgments}

Authors are thankful to the anonymous referee for useful comments which 
improved the scientific content and presentation of the paper.
This publication makes use of data from the 2MASS (a joint project of the 
University of Massachusetts and the Infrared Processing and Analysis Center
/California Institute of Technology, funded by the National Aeronautics 
and Space Administration and the National Science Foundation). 
This research has made use of the WEBDA database, operated  at the
Institute for Astronomy of the University of Vienna, as well as has used 
the images from the Digital Sky Survey  (DSS), which
was produced  at the  Space Telescope Science  Institute under  the US
Government grant  NAG W-2166. We have also used  NASA's Astrophysics Data System
and {\small IRAF}, distributed   by  National   Optical  Astronomy
Observatories,  USA.  

{}

\begin{table}
\begin{small}
\caption{Observed polarized and unpolarized standard stars}
\label{tab_standards}
\begin{tabular}{lllll}\hline \hline
 & P$\pm\epsilon_{P}$($\%$) &$\theta\pm\epsilon_{\theta}$($\degr$) &P$\pm\epsilon_{P}$($\%$)&  $\theta\pm\epsilon_{\theta}$($\degr$)\\
\multicolumn{5}{c}{\hspace{0.5cm}Our work\hspace{2.4cm}Schmidt et al. (2002)}\\
\hline
\multicolumn{5}{c}{Polarized standard stars}\\
\hline
\multicolumn{5}{c}{2008 November 8 $\&$ 9}\\
\hline      
\multicolumn{5}{c}{{\bf BD+59$\degr$389}}\\
        $B$       &        6.32$\pm$ 0.22 & 97.9$\pm$1.0 & 6.34 $\pm$0.04 & 98.14 $\pm$0.16  \\
        $V$       &         6.86$\pm$ 0.16 & 97.3$\pm$0.7 & 6.70 $\pm$0.01 & 98.09 $\pm$0.07  \\
        $R_{c}$ &     6.43$\pm$ 0.13 & 96.0$\pm$0.6 & 6.43 $\pm$0.02 & 98.14$\pm$0.10   \\
        $I_c$&       5.86$\pm$ 0.11 & 97.0$\pm$0.5 & 5.80 $\pm$0.02 & 98.26$\pm$0.11   \\

\multicolumn{5}{c}{{\bf BD+64$\degr$ 106}}\\
         $B$ & 5.73$\pm$ 0.21& 97.3$\pm$1.0& 5.51 $\pm$0.09 & 97.15$\pm$0.47   \\
         $V$ &        5.68$\pm$ 0.19& 96.7$\pm$1.0& 5.69 $\pm$0.04 & 96.63$\pm$0.18   \\
         $R_c$&       5.12$\pm$ 0.17& 97.4$\pm$0.9& 5.15 $\pm$0.10 & 96.74$\pm$0.54   \\
         $I_c$&       4.66$\pm$ 0.21& 97.8$\pm$1.3& 4.70 $\pm$0.05 & 96.89$\pm$0.32   \\

\multicolumn{5}{c}{{\bf BD+64$\degr$ 106}}\\
         $B$ & 5.65$\pm$ 0.22 & 97.4$\pm$1.1& 5.51 $\pm$0.09 & 97.15$\pm$0.47   \\
         $V$ &         5.49$\pm$ 0.20 & 96.9$\pm$1.0& 5.69 $\pm$0.04 & 96.63$\pm$0.18   \\
         $R_c$&        5.33$\pm$ 0.18 & 95.8$\pm$0.9& 5.15 $\pm$0.10 & 96.74$\pm$0.54   \\
        $I_c$ &        4.48$\pm$ 0.22 & 98.7$\pm$1.4& 4.70 $\pm$0.05 & 96.89$\pm$0.32   \\

\multicolumn{5}{c}{{\bf HD204827}}\\
       $B$ &         5.68$\pm$ 0.23& 58.0$\pm$1.2& 5.65 $\pm$0.02 & 58.20$\pm$0.11   \\
       $V$ &          5.47$\pm$ 0.17& 59.1$\pm$0.8& 5.32 $\pm$0.01 & 58.73$\pm$0.08   \\
       $R_c$&         5.19$\pm$ 0.14& 61.5$\pm$0.8& 4.89 $\pm$0.03 & 59.10$\pm$0.17   \\
       $I_c$&         4.02$\pm$ 0.18& 58.5$\pm$1.2& 4.19 $\pm$0.03 & 59.94$\pm$0.20   \\
\hline
\multicolumn{5}{c}{2009 November 23}\\
\hline
\multicolumn{5}{c}{{\bf HD19820}}\\
                 $B$     & 4.49 $\pm$ 0.11 & 114.9 $\pm$ 0.7 &  4.699 $\pm$ 0.036 &  115.70 $\pm$ 0.22 \\
                 $V$      &  4.89 $\pm$ 0.09 & 114.2 $\pm$ 0.5 &  4.787 $\pm$ 0.028 &  114.93 $\pm$ 0.17 \\
                 $R_c$  &  4.49 $\pm$ 0.09 & 115.5 $\pm$ 0.6 &  4.526 $\pm$ 0.025 &  114.46 $\pm$ 0.16 \\
                 $I_c$  &  4.06 $\pm$ 0.16 & 115.3 $\pm$ 1.0 &  4.081 $\pm$ 0.024 &  114.48 $\pm$ 0.17 \\

\multicolumn{5}{c}{{\bf HD25443}}\\
                  $B$       &  5.19 $\pm$  0.09 & 134.6 $\pm$ 0.5  &  5.232 $\pm$  0.092 & 134.28 $\pm$ 0.51 \\
                  $V$       &  5.04 $\pm$  0.07 & 136.0 $\pm$ 0.4  &  5.127 $\pm$  0.061 & 134.23 $\pm$ 0.34 \\
                 $I_c$   &  4.19 $\pm$  0.09 & 134.8 $\pm$ 0.6  &  4.249 $\pm$  0.041 & 134.21 $\pm$ 0.28 \\

\multicolumn{5}{c}{{\bf BD+64$\degr$106}}\\ 
                  $B$      &  5.49 $\pm$  0.17 & 98.0 $\pm$  0.9  &  5.506 $\pm$  0.090 &  97.15 $\pm$ 0.47 \\
                 $R_c$   &  5.41 $\pm$ 0.11 & 96.1 $\pm$  0.6  &  5.150 $\pm$  0.098 &  96.74 $\pm$  0.54 \\
                 $I_c$   &  4.50 $\pm$ 0.14 & 96.6 $\pm$  0.9  &  4.696 $\pm$  0.052 &  96.89 $\pm$  0.32 \\
\hline
\multicolumn{5}{c}{2009 December 24}\\
\hline
\multicolumn{5}{c}{{\bf HD19820}}\\
                  $B$      & 4.72 $\pm$ 0.11 & 115.6 $\pm$ 0.7  &   4.699 $\pm$ 0.036 &  115.70 $\pm$ 0.22 \\
                  $V$      & 4.79  $\pm$ 0.08 & 115.1 $\pm$ 0.5 &   4.787 $\pm$ 0.028 &  114.93 $\pm$ 0.17 \\
                  $R_c$    & 4.51 $\pm$ 0.07  & 114.6 $\pm$ 0.4 &   4.526 $\pm$ 0.025 &  114.46 $\pm$ 0.16 \\
                  $I_c$    & 3.97 $\pm$ 0.09  & 115.7 $\pm$ 0.6 &   4.081 $\pm$ 0.024 &  114.48 $\pm$ 0.17 \\

\multicolumn{5}{c}{{\bf HD25443}}\\
                  $B$      & 5.12 $\pm$ 0.09 & 134.0 $\pm$ 0.5  &   5.232 $\pm$  0.092 & 134.28 $\pm$ 0.51 \\
                  $V$      & 5.27 $\pm$ 0.09 & 134.8 $\pm$ 0.5  &   5.127 $\pm$  0.061 & 134.23 $\pm$ 0.34 \\
                  $R_c$    & 5.00 $\pm$ 0.08 & 134.5 $\pm$ 0.4  &   4.734 $\pm$  0.045 & 133.65 $\pm$ 0.28 \\
                  $I_c$    & 4.19 $\pm$ 0.07 & 136.0 $\pm$ 0.5  &   4.249 $\pm$  0.041 & 134.21 $\pm$ 0.28 \\
\hline
\multicolumn{5}{c}{Unpolarized standard stars}\\ 
\hline
\multicolumn{5}{c}{\hspace{0.1cm}{\bf BD+32$\degr$3739}\hspace{2.0cm}{\bf HD212311}}\\
                         &       q (\%)  & u (\%)  &  q (\%) & u (\%) \\
\multicolumn{5}{c}{This wrok}\\
\hline 
                  $B$     &   0.089   &    -0.093 &    0.138  &   0.001  \\
                  $V$     &   0.171   &    -0.058 &   -0.184  &  -0.043  \\
                  $R_c$   &  -0.072   &    -0.078 &    0.042  &   0.016  \\
                  $I_c$   &   0.058   &    -0.010 &   -0.162 &   -0.139  \\
\hline \hline
\end{tabular}
\end{small}
\end{table}

\begin{table}
\begin{small}
\caption{Observed $B$, $V$, $(RI)_{c}$ polarization and polarization angles for stars towards NGC 1893}\label{result_1893}
\begin{tabular}{cccccccccc}\hline \hline
Id$^{\dagger}$ &  V (mag)$^{\ddagger}$ & $P_{B}\pm\epsilon(\%)$& $\theta_{B}\pm\epsilon$($\degr$)& $P_{V}\pm\epsilon(\%)$& $\theta_{V}\pm\epsilon$($\degr$)& $P_{Rc}\pm\epsilon(\%)$
& $\theta_{Rc}\pm\epsilon$($\degr$)& $P_{Ic}\pm\epsilon(\%)$& $\theta_{Ic}\pm\epsilon$($\degr$) \\  
(1)&(2)&(3)&(4)&(5)&(6)&(7)&(8)&(9)&(10) \\
\hline 
   1                  &  10.19  &   2.35 $\pm$ 0.14   &   160.3 $\pm$ 1.6   &   2.52 $\pm$ 0.14   &   159.5 $\pm$ 1.5   &   2.44 $\pm$ 0.14   &   154.9$\pm$  1.5   &   1.93 $\pm$ 0.18  &    155.6 $\pm$ 2.5   \\
   2                  &  11.29  &   2.90 $\pm$ 0.23   &   157.0 $\pm$ 2.1   &   3.15 $\pm$ 0.24   &   160.8 $\pm$ 1.9   &   3.06 $\pm$ 0.22   &   156.6$\pm$  2.0   &   2.38 $\pm$ 0.29  &    164.4 $\pm$ 3.4   \\
   3                  &  11.16  &   2.31 $\pm$ 0.16   &   157.3 $\pm$ 1.9   &   2.30 $\pm$ 0.15   &   163.1 $\pm$ 1.8   &   2.27 $\pm$ 0.15   &   162.1$\pm$  1.9   &   2.28 $\pm$ 0.18  &    159.1 $\pm$ 2.2   \\
   5                  &  13.41  &   2.22 $\pm$ 0.46   &   156.8 $\pm$ 5.6   &   2.79 $\pm$ 0.42   &   158.0 $\pm$ 4.1   &   2.30 $\pm$ 0.41   &   155.5$\pm$  5.0   &   1.95 $\pm$ 0.47  &    151.7 $\pm$ 6.6   \\
   9                  &  12.71  &   2.54 $\pm$ 0.34   &   160.8 $\pm$ 3.7   &   2.76 $\pm$ 0.31   &   165.7 $\pm$ 3.0   &   2.87 $\pm$ 0.29   &   162.9$\pm$  2.8   &   2.03 $\pm$ 0.32  &    156.2 $\pm$ 4.3   \\
  10                  &  13.69  &   2.15 $\pm$ 0.24   &   161.9 $\pm$ 3.1   &   2.03 $\pm$ 0.23   &   157.2 $\pm$ 3.0   &   2.28 $\pm$ 0.23   &   162.9$\pm$  2.7   &   2.10 $\pm$ 0.26  &    158.6 $\pm$ 3.3   \\
  13                  &  12.46  &   2.28 $\pm$ 0.52   &   153.3 $\pm$ 6.3   &   1.95 $\pm$ 0.46   &   157.4 $\pm$ 6.3   &   3.01 $\pm$ 0.44   &   157.9$\pm$  4.0   &   2.38 $\pm$ 0.45  &    161.8 $\pm$ 5.3   \\
  15                  &  13.02  &   2.66 $\pm$ 0.42   &   156.4 $\pm$ 4.2   &   3.08 $\pm$ 0.48   &   165.8 $\pm$ 3.7   &   2.43 $\pm$ 0.44   &   161.9$\pm$  4.9   &   2.90 $\pm$ 0.54  &    162.6 $\pm$ 5.2   \\
  33                  &  12.27  &   2.11 $\pm$ 0.35   &   161.9 $\pm$ 4.3   &   2.52 $\pm$ 0.38   &   168.0 $\pm$ 3.5   &   2.04 $\pm$ 0.34   &   159.2$\pm$  4.5   &   2.15 $\pm$ 0.44  &    160.2 $\pm$ 5.6   \\
  48                  &  11.01  &   1.83 $\pm$ 0.27   &   166.0 $\pm$ 4.0   &   1.81 $\pm$ 0.19   &   165.6 $\pm$ 2.7   &   1.32 $\pm$ 0.18   &   164.2$\pm$  3.7   &   1.59 $\pm$ 0.17  &    165.2 $\pm$ 2.9   \\
  59                  &  12.10  &   2.32 $\pm$ 0.38   &   156.8 $\pm$ 4.4   &   2.79 $\pm$ 0.40   &   159.6 $\pm$ 3.6   &   2.95 $\pm$ 0.36   &   155.4$\pm$  3.4   &   2.38 $\pm$ 0.46  &    161.6 $\pm$ 5.3   \\
  62                  &  12.69  &   2.06 $\pm$ 0.31   &   162.8 $\pm$ 4.2   &   2.59 $\pm$ 0.29   &   159.3 $\pm$ 3.0   &   2.44 $\pm$ 0.27   &   158.0$\pm$  3.1   &   1.91 $\pm$ 0.29  &    163.9 $\pm$ 4.2   \\
  72                  &  12.53  &   1.40 $\pm$ 0.43   &   172.4 $\pm$ 8.2   &   1.16 $\pm$ 0.31   &   174.7 $\pm$ 7.2   &   1.87 $\pm$ 0.34   &   162.0$\pm$  5.0   &   1.81 $\pm$ 0.39  &    162.2 $\pm$ 5.8   \\
  73                  &  10.77  &   1.66 $\pm$ 0.18   &   169.2 $\pm$ 2.7   &   1.89 $\pm$ 0.19   &   169.2 $\pm$ 2.3   &   1.72 $\pm$ 0.17   &   167.8$\pm$  2.7   &   1.36 $\pm$ 0.23  &    173.9 $\pm$ 4.5   \\
  82                  &  12.94  &   2.67 $\pm$ 0.52   &   154.5 $\pm$ 5.2   &   3.25 $\pm$ 0.48   &   153.8 $\pm$ 4.1   &   3.56 $\pm$ 0.44   &   150.4$\pm$  3.4   &   2.82 $\pm$ 0.53  &    165.3 $\pm$ 5.2   \\
  92                  &  11.72  &   1.27 $\pm$ 0.26   &   169.5 $\pm$ 5.5   &   1.54 $\pm$ 0.25   &   162.3 $\pm$ 4.2   &   1.47 $\pm$ 0.30   &   160.5$\pm$  5.5   &   1.28 $\pm$ 0.34  &    158.4 $\pm$ 7.1   \\
  98                  &  13.28  &   3.67 $\pm$ 0.56   &   158.9 $\pm$ 4.3   &   3.40 $\pm$ 0.51   &   157.3 $\pm$ 4.1   &   2.96 $\pm$ 0.58   &   157.3$\pm$  5.4   &   3.05 $\pm$ 0.60  &    165.1 $\pm$ 5.4   \\
 106                  &  12.32  &   2.97 $\pm$ 0.29   &   160.3 $\pm$ 2.7   &   2.66 $\pm$ 0.26   &   159.0 $\pm$ 2.7   &   2.64 $\pm$ 0.25   &   159.4$\pm$  2.6   &   2.54 $\pm$ 0.28  &    162.7 $\pm$ 3.0   \\
 108                  &  13.88  &   2.55 $\pm$ 0.59   &   154.8 $\pm$ 6.4   &   3.67 $\pm$ 0.54   &   159.3 $\pm$ 4.0   &   2.94 $\pm$ 0.52   &   165.9$\pm$  4.9   &   2.26 $\pm$ 0.58  &    157.9 $\pm$ 7.0   \\
 115                  &  13.66  &   2.72 $\pm$ 0.48   &   162.4 $\pm$ 4.9   &   2.80 $\pm$ 0.43   &   160.3 $\pm$ 4.1   &   2.98 $\pm$ 0.40   &   163.0$\pm$  3.8   &   2.36 $\pm$ 0.43  &    158.3 $\pm$ 4.9   \\
 123            &  13.23$^{a}$  &   3.65 $\pm$ 0.42   &   163.6 $\pm$ 3.2   &   3.57 $\pm$ 0.40   &   160.9 $\pm$ 3.0   &   4.04 $\pm$ 0.39   &   157.3$\pm$  2.7   &   2.31 $\pm$ 0.42  &    158.1 $\pm$ 5.0   \\
 134                  &  13.09  &   2.91 $\pm$ 0.40   &   160.4 $\pm$ 3.8   &   3.38 $\pm$ 0.38   &   167.1 $\pm$ 3.0   &   2.74 $\pm$ 0.37   &   155.2$\pm$  3.7   &   3.07 $\pm$ 0.41  &    164.6 $\pm$ 3.7   \\
 139                  &  11.98  &   2.20 $\pm$ 0.27   &   161.4 $\pm$ 3.3   &   2.76 $\pm$ 0.25   &   160.4 $\pm$ 2.5   &   2.82 $\pm$ 0.31   &   160.5$\pm$  3.1   &   2.46 $\pm$ 0.42  &    160.6 $\pm$ 4.7   \\
 140                  &  12.39  &   2.65 $\pm$ 0.33   &   165.1 $\pm$ 3.4   &   2.48 $\pm$ 0.30   &   158.5 $\pm$ 3.3   &   2.87 $\pm$ 0.37   &   157.9$\pm$  3.6   &   2.30 $\pm$ 0.49  &    154.9 $\pm$ 5.8   \\
 141                  &  11.45  &   2.49 $\pm$ 0.23   &   164.3 $\pm$ 2.4   &   2.53 $\pm$ 0.25   &   160.1 $\pm$ 2.5   &   2.40 $\pm$ 0.23   &   157.9$\pm$  2.7   &   2.48 $\pm$ 0.31  &    162.6 $\pm$ 3.5   \\
 149                  &  10.14  &   2.69 $\pm$ 0.13   &   157.2 $\pm$ 1.3   &   2.68 $\pm$ 0.11   &   149.8 $\pm$ 1.1   &   2.87 $\pm$ 0.12   &   149.9$\pm$  1.2   &   2.33 $\pm$ 0.15  &    144.4 $\pm$ 1.7   \\
 168                  &  12.25  &   2.12 $\pm$ 0.33   &   160.1 $\pm$ 4.3   &   2.36 $\pm$ 0.32   &   154.3 $\pm$ 3.7   &   2.01 $\pm$ 0.40   &   146.8$\pm$  5.4   &   2.36 $\pm$ 0.44  &    152.6 $\pm$ 5.1   \\
 196             &  12.33$^{a}$  &   1.82 $\pm$ 0.33   &   164.4 $\pm$ 4.9   &   2.45 $\pm$ 0.29   &   159.7 $\pm$ 3.1   &   2.17 $\pm$ 0.25   &   164.5$\pm$  3.1   &   2.07 $\pm$ 0.26  &    167.8 $\pm$ 3.4   \\
 197                  &  11.87  &   2.33 $\pm$ 0.35   &   162.4 $\pm$ 4.1   &   2.13 $\pm$ 0.22   &   161.3 $\pm$ 2.7   &   1.99 $\pm$ 0.16   &   163.3$\pm$  2.2   &   2.00 $\pm$ 0.14  &    164.7 $\pm$ 1.9   \\
 199                   &  13.40  &   1.82 $\pm$ 0.46   &   157.3 $\pm$ 6.8   &   3.77 $\pm$ 0.40   &   154.9 $\pm$ 2.9   &   2.59 $\pm$ 0.37   &   146.2$\pm$  3.9   &   2.16 $\pm$ 0.36  &    148.2 $\pm$ 4.6   \\
 209             &  12.63$^{a}$  &   2.72 $\pm$ 0.29   &   165.7 $\pm$ 3.0   &   1.80 $\pm$ 0.29   &   168.2 $\pm$ 4.2   &   1.84 $\pm$ 0.29   &   167.9$\pm$  4.3   &   1.79 $\pm$ 0.34  &    163.4 $\pm$ 5.2   \\
 213                  &  14.52  &   2.98 $\pm$ 0.78   &   143.6 $\pm$ 7.2   &   3.99 $\pm$ 0.69   &   165.2 $\pm$ 4.7   &   2.72 $\pm$ 0.62   &   159.7$\pm$  6.3   &   2.84 $\pm$ 0.64  &    173.6 $\pm$ 6.2   \\
 215                  &  14.23  &   3.42 $\pm$ 0.69   &   156.8 $\pm$ 5.6   &   2.72 $\pm$ 0.63   &   164.8 $\pm$ 6.1   &   2.55 $\pm$ 0.58   &   157.7$\pm$  6.3   &   2.02 $\pm$ 0.62  &    154.4 $\pm$ 8.4   \\
 219                  &  13.97  &   3.82 $\pm$ 0.63   &   156.7 $\pm$ 4.6   &   3.38 $\pm$ 0.56   &   161.8 $\pm$ 4.4   &   4.12 $\pm$ 0.51   &   155.0$\pm$  3.5   &   2.67 $\pm$ 0.55  &    159.5 $\pm$ 5.6   \\
 222                  &  14.16  &   2.32 $\pm$ 0.65   &   169.0 $\pm$ 7.6   &   2.61 $\pm$ 0.60   &   154.7 $\pm$ 6.3   &   2.33 $\pm$ 0.59   &   164.4$\pm$  6.9   &   2.31 $\pm$ 0.66  &    164.0 $\pm$ 7.9   \\
 228                  &  12.50  &   2.12 $\pm$ 0.35   &   162.0 $\pm$ 4.5   &   3.13 $\pm$ 0.32   &   162.6 $\pm$ 2.8   &   2.71 $\pm$ 0.39   &   155.7$\pm$  4.0   &   2.37 $\pm$ 0.50  &    158.8 $\pm$ 5.9   \\
 301             &  10.94$^{a}$  &   3.16 $\pm$ 0.13   &   160.0 $\pm$ 1.2   &   3.08 $\pm$ 0.13   &   160.9 $\pm$ 1.2   &   2.92 $\pm$ 0.13   &   161.9$\pm$  1.3   &   2.96 $\pm$ 0.16  &    162.5 $\pm$ 1.4   \\
 307              &  11.68$^{a}$ &   3.14 $\pm$ 0.19   &   159.7 $\pm$ 1.7   &   3.48 $\pm$ 0.19   &   159.9 $\pm$ 1.5   &   3.33 $\pm$ 0.19   &   159.3$\pm$  1.6   &   3.02 $\pm$ 0.22  &    158.8 $\pm$ 2.0   \\
 310             &  12.59$^{a}$  &   2.33 $\pm$ 0.42   &   160.4 $\pm$ 4.9   &   2.00 $\pm$ 0.29   &   167.6 $\pm$ 3.8   &   1.71 $\pm$ 0.23   &   166.9$\pm$  3.6   &   1.53 $\pm$ 0.21  &    163.6 $\pm$ 3.7   \\
 320             &  12.43$^{a}$  &   2.59 $\pm$ 0.29   &   164.7 $\pm$ 3.1   &   2.88 $\pm$ 0.29   &   161.6 $\pm$ 2.8   &   2.65 $\pm$ 0.36   &   165.8$\pm$  3.7   &   2.38 $\pm$ 0.46  &    160.6 $\pm$ 5.3   \\
 327             &  11.62$^{a}$  &   1.25 $\pm$ 0.22   &   161.7 $\pm$ 4.7   &   1.01 $\pm$ 0.21   &   164.3 $\pm$ 5.3   &   1.06 $\pm$ 0.26   &   156.0$\pm$  6.5   &   1.05 $\pm$ 0.34  &    168.4 $\pm$ 8.5   \\
 329             &  11.95$^{a}$  &   1.20 $\pm$ 0.26   &   152.7 $\pm$ 5.7   &   0.84 $\pm$ 0.24   &   160.5 $\pm$ 7.4   &   1.03 $\pm$ 0.32   &   148.4$\pm$  8.0   &   1.49 $\pm$ 0.42  &    164.5 $\pm$ 7.6   \\
 355             &  13.46$^{a}$  &   1.61 $\pm$ 0.53   &   154.8 $\pm$ 8.8   &   1.94 $\pm$ 0.47   &   168.2 $\pm$ 6.6   &   2.01 $\pm$ 0.53   &   168.7$\pm$  7.2   &   2.01 $\pm$ 0.64  &    170.8 $\pm$ 8.6   \\
 356             &  14.35$^{a}$  &   2.57 $\pm$ 0.80   &   160.7 $\pm$ 8.6   &   2.01 $\pm$ 0.67   &   164.0 $\pm$ 8.7   &   3.23 $\pm$ 0.58   &   157.3$\pm$  5.0   &   1.85 $\pm$ 0.60  &    150.6 $\pm$ 8.7   \\
\hline \hline                                                                                          
\end{tabular} \\
$^{\dagger}$: Cuffey and Shapley (1937)\\
$^{\ddagger}$: Sharma et al. (2007)\\
$^a$: Massey et al. (1995)\\
\end{small}                                                                                         
\end{table}


\begin{table}
\caption{Polarimetric results for 14 stars observed towards NGC 2281. 
}\label{result_2281}
\begin{tabular}{ccccccccccc}\hline \hline
Id.$^{\dagger}$ & $\alpha (\degr)$          & $\delta (\degr)$    & $V$  & $(B-V)$$^{*}$ & $(U-B)$$^{*}$ & $P_{V}$ &  $\theta_{V}$ & M$_{p}^{\dagger\dagger}$ & $M_{p}^{\dagger\dagger\dagger}$ \\
              & (J2000) & (J2000) &  (mag)$^{*}$   &   &    & $(\%) $  & $(\degr)$  &  & \\
(1) & (2) & (3) & (4) & (5) & (6) & (7) & (8) & (9) & (10) \\
\hline
 49 & 102.03038 & 41.077494  & 10.56 &  0.24  &  0.16 &  1.07$\pm$ 0.17 & 16.16 $\pm$ 4.28 & M   & M (99)  \\
 53 & 102.06064 & 41.112788  & 11.02 &  0.33  &  0.13 &  0.91$\pm$ 0.22 & 19.52 $\pm$ 6.14 & M   & M (99)  \\
 55 & 102.06337 & 41.073019  &  8.89 &  0.98  &  0.69 &  1.00$\pm$ 0.08 & 13.77 $\pm$ 2.10 & NM   & M (99)  \\
 56 & 102.06596 & 41.095871  & 11.43 &  1.24  &  1.19 &  1.04$\pm$ 0.25 &  6.23 $\pm$ 6.39 & NM   & NM (00)  \\
 57 & 102.07032 & 41.130267  & 10.61 &  0.23  &  0.15 &  0.90$\pm$ 0.18 &  6.39 $\pm$ 5.05 & M   & M (99)  \\
 58 & 102.07208 & 41.083118  &  9.45 &  0.12  &  0.06 &  0.73$\pm$ 0.10 &  9.65 $\pm$ 3.53 & M   & M (99)  \\
 60 & 102.08048 & 41.041193  & 12.71 &  0.56  &  0.07 &  1.41$\pm$ 0.45 & 12.29 $\pm$ 8.47 & M   & M (99)  \\
 71 & 102.11072 & 41.024925  & 11.16 &  0.53  &  0.05 &  0.78$\pm$ 0.23 & 24.01 $\pm$ 7.43 & M   & NM (00)  \\
 74 & 102.1166  & 41.081101  &  9.54 &  1.10  &  1.03 &  0.56$\pm$ 0.11 & 22.08 $\pm$ 4.62 & NM   & NM (00)  \\
 78 & 102.13344 & 41.108674  & 10.63 &  0.22  &  0.15 &  0.60$\pm$ 0.18 & 23.54 $\pm$ 7.22 & M   & M (99)  \\
 79 & 102.14759 & 41.090775  & 10.30 &  0.17  &  0.11 &  0.67$\pm$ 0.15 & 12.83 $\pm$ 5.60 & M   & M (99)  \\
 82 & 102.15634 & 41.10563   & 10.08 &  0.14  &  0.10 &  0.85$\pm$ 0.14 & 18.06 $\pm$ 4.11 & M   & M (99)  \\
 86 & 102.17038 & 41.066299  &  8.62 &  0.22  &  0.15 &  0.93$\pm$ 0.07 & 19.43 $\pm$ 1.92 & M   & M (99)  \\
 91 & 102.18262 & 41.086379  & 11.41 &  0.35  &  0.12 &  0.94$\pm$ 0.25 & 25.04 $\pm$ 6.97 & M   & M (99)  \\

\hline                                           
\hline                                           
\end{tabular}\\
$^{\dagger}$:  Vasilevskis \& Balz (1959)\\
$^{*}$: Pesch (1961) except star no \# 60 for which $V$ magnitude, $(B-V)$ and $(U-B)$ 
colours are taken from Yoshizawa (1978)\\
$^{\dagger\dagger}$: Present Work\\
$^{\dagger\dagger\dagger}$: Vasilevskis \& Balz (1959)\\
\end{table}

\begin{table}
\caption{Polarimetric results on 27 stars observed towards NGC 1664.}\label{result_1664}
\begin{tabular}{cccccccccc}\hline \hline
Id.$^{\dagger}$&$\alpha (\degr)$          &$\delta (\degr)$            & $V$ & $(B-V)$  & $(U-B)$  & $P_{V}$ &  $\theta_{V}$ & M$_{p}$ $^{\dagger\dagger}$ & M$_{p}$ $^{\dagger\dagger\dagger}$ \\
               &(J2000)& (J2000)&  (mag)     &   &   & $(\%)$   & $(\degr)$  &  &  \\
(1) & (2) & (3) & (4) & (5) & (6) & (7) & (8) & (9) & (10) \\
\hline
  1 & 72.783342 & 43.665161  & 13.50$^{a}$    &  1.61$^{a}$  &  1.68$^{a}$    &   2.44 $\pm$  0.35 &  165 $\pm$    4 & NM    &  -      \\
  4 & 72.769500 & 43.671624  & 11.75$^{b}$    &  0.32$^{b}$  &  0.25$^{b}$    &   1.41 $\pm$  0.15 &  178 $\pm$    3 & M    &  M (65)   \\
  7 & 72.824534 & 43.671985  & 12.25$^{a}$    &  0.30$^{a}$  &  0.24$^{a}$    &   1.14 $\pm$  0.18 &  165 $\pm$    4 & M    &  M (89)   \\
  9 & 72.814747 & 43.680135  &   -            &   -         &   -             &   2.75 $\pm$  0.14 &  164 $\pm$    1 & NM    &   -      \\
 12 & 72.749866 & 43.684928  & 13.01$^{a}$    &  0.24$^{a}$  &  0.31$^{a}$    &   1.25 $\pm$  0.27 &  170 $\pm$    6 & M    &  M (92)   \\
 13 & 72.760895 & 43.676324  & 14.36$^{b}$    &  1.09$^{b}$  &  0.71$^{b}$    &   2.59 $\pm$  0.49 &  163 $\pm$    5 & NM   &  -      \\
 14 & 72.750261 & 43.676335  & 12.79$^{a}$    &  0.51$^{a}$  &  0.28$^{a}$    &   1.25 $\pm$  0.26 &  170 $\pm$    5 & M    &  M (91)   \\
 21 & 72.771228 & 43.648455  & 11.81$^{b}$    &  0.29$^{b}$  &  0.26$^{b}$    &   1.39 $\pm$  0.15 &  173 $\pm$    3 & M    &  M (82)   \\
 22 & 72.779262 & 43.640397  & 14.00$^{a}$    &  0.54$^{a}$  &  0.38$^{a}$    &   2.33 $\pm$  0.42 &  171 $\pm$    5 & NM    &  M (86)   \\
 27 & 72.816757 & 43.656291  & 13.87$^{b}$    &  0.51$^{b}$  &  0.21$^{b}$    &   2.10 $\pm$  0.39 &  175 $\pm$    5 & NM    &  NM (35)   \\
 35 & 72.804006 & 43.700468  & 12.50$^{c}$    &  0.37$^{c}$  &  0.25$^{c}$    &   1.15 $\pm$  0.21 &  170 $\pm$    5 & M    &  M (91)   \\
 37 & 72.790902 & 43.708346  & 10.93$^{b}$    &  0.34$^{b}$  &  0.34$^{b}$    &   1.24 $\pm$  0.10 &  174 $\pm$    2 & M    &  M (90)   \\
 47 & 72.754574 & 43.703638  &   -            &  -          &   -             &   3.18 $\pm$  0.54 &  160 $\pm$    5 & NM    &  -      \\
 55 & 72.724155 & 43.662572  & 11.06$^{b}$    &  1.23$^{b}$  &  1.00$^{b}$    &   1.71 $\pm$  0.11 &  169 $\pm$    2 & M    &  M (71)   \\
 56 & 72.737611 & 43.657761  & 12.66$^{b}$    &  0.35$^{b}$  &  0.26$^{b}$    &   1.40 $\pm$  0.23 &  177 $\pm$    4 & M    &  M (91)   \\
 67 & 72.857380 & 43.672892  &  -             &  -          &   -             &   1.21 $\pm$  0.38 &  171 $\pm$    8 & M    &  M (92)   \\
 68 & 72.849878 & 43.675125  &  -             &  -          &   -             &   1.57 $\pm$  0.27 &  168 $\pm$    5 & M    &  M (87)   \\
 75 & 72.832061 & 43.704486  & 11.25$^{a}$    &  1.06$^{a}$  &  0.83$^{a}$    &   1.47 $\pm$  0.13 &  173 $\pm$    2 & M    &  M (92)   \\
 76 & 72.823740 & 43.720916  &  -             &  -          &   -             &   1.53 $\pm$  0.36 &  177 $\pm$    6 & M    &  M (87)   \\
 77 & 72.814271 & 43.709793  & 12.84$^{c}$    &  0.49$^{c}$  &  0.25$^{c}$    &   1.40 $\pm$  0.25 &  174 $\pm$    5 & M    &  M (65)   \\
 79 & 72.765348 & 43.727229  & 13.75$^{a}$    &  0.54$^{a}$  &  0.50$^{a}$    &   2.06 $\pm$  0.38 &  165 $\pm$    5 & NM   &  -      \\
 87 & 72.700196 & 43.654016  &   -            &  -          &   -             &   4.09 $\pm$  0.80 &  171 $\pm$    5 & NM    &  -      \\
 90 & 72.712849 & 43.643313  & 13.03$^{a}$    &  0.51$^{a}$  &  0.30$^{a}$    &   1.47 $\pm$  0.29 &  170 $\pm$    5 & M    &  NM (17)   \\
113 & 72.735822 & 43.721862  & 12.97$^{a}$    &  0.53$^{a}$  &  0.27$^{a}$    &   1.45 $\pm$  0.29 &  175 $\pm$    5 & M    &  M (84)   \\
154 & 72.780622 & 43.746898  &    -           &  -          &    -             &   1.26 $\pm$  0.31 &  172 $\pm$    6 & M    &  -      \\
158 & 72.719637 & 43.739690  & 14.79$^{a}$    &  0.73$^{a}$  &  0.13$^{a}$    &   1.78 $\pm$  0.50 &  166 $\pm$    7 & M    &  -      \\
164 & 72.691065 & 43.720635  &    -           &   -         &   -             &   0.97 $\pm$  0.27 &   26 $\pm$    7 & NM    &  NM (00)   \\
\hline                                           
\hline                                           
\end{tabular}\\ 
$^{\dagger}$:  Larsson-Leander (1957) \\  
$^{a}$: Hoag et al. (1961) Publ. Us. Nav. Obs. XVII part VII, 347 \\
$^{b}$: Purgathofer (1964)\\
$^{c}$: Hoag et al. (1961)\\ 
$^{\dagger\dagger}$: Present Work\\
$^{\dagger\dagger\dagger}$: Dias et al. (2006)\\

\end{table}

\begin{table}
\caption{Polarimetric results on 15 stars observed towards NGC 1960.}\label{result_1960}
\begin{tabular}{ccccccccccc}\hline \hline
Id.$^{\dagger}$ & $\alpha (\degr)$       & $\delta (\degr)$      & $V$ & $(B-V)$$^{*}$ & $ (U-B)$$^{*}$ & $P_{V}$ &  $\theta_{V}$ & M$_{p}$ $^{\dagger\dagger}$ &M$_{p}$ $^{\dagger\dagger\dagger}$ & \\
               &(J2000)& (J2000)&  (mag)$^{*}$    &   &   &  $(\%)$   & $(\degr)$ & & \\
(1) & (2) & (3) & (4) & (5) & (6) & (7) & (8) & (9) & (10) \\
\hline
 13 & 84.063492 & 34.120250  & 10.78        &  0.12         &   -0.26        & 1.12 $\pm$ 0.17 & 160  $\pm$  4  & M & M (94)  \\  
 16 & 84.065508 & 34.143746  &  8.86$^{a}$  &  -0.00$^{a}$   &  -0.66 $^{a}$   & 1.17 $\pm$ 0.07 & 159  $\pm$  2  & M & M (94)   \\ 
 17 & 84.060257 & 34.139917  & 12.40        &  0.25         &   0.18        & 1.16 $\pm$ 0.35 & 162  $\pm$  8  & M & M (94)  \\  
 23 & 84.095740 & 34.175955  &  8.96$^{a}$  &  0.01$^{a}$   &    -0.68 $^{a}$   & 1.12 $\pm$ 0.07 & 162  $\pm$  2  & M & M (94)  \\  
 33 & 84.118160 & 34.122944  & 11.85       &   0.12        &    -0.07        & 1.15 $\pm$ 0.27 & 155  $\pm$  6   & M & M (94)  \\  
 38 & 84.099176 & 34.099402  &  9.92$^{a}$  &  0.05$^{a}$   &    -0.49$^{a}$   & 1.17 $\pm$ 0.11 & 156  $\pm$  2  & M & M (94)  \\  
 41 & 84.083418 & 34.103419  & 12.37       &   0.19        &    0.14        & 1.28 $\pm$ 0.35 & 160  $\pm$  7   & M & M (94)  \\  
 44 & 84.047893 & 34.118557  & 11.35       &   0.09        &    -0.36        & 1.38 $\pm$ 0.22 & 163  $\pm$  4   & M & M (94)  \\  
 55 & 84.081130 & 34.204218  & 11.61       &   0.10        &    -0.31        & 1.44 $\pm$ 0.24 & 147  $\pm$  5   & NM & M (94)  \\  
 56 & 84.087224 & 34.211974  & 12.39       &   0.46        &    0.35        & 1.15 $\pm$ 0.26 & 157  $\pm$  6   & M & M (93)  \\  
 61 & 84.132976 & 34.179480  &  9.14$^{a}$ &   0.01$^{a}$  &     -0.66$^{a}$  & 0.81 $\pm$ 0.08 & 155  $\pm$  2   & NM & M (94)  \\  
 77 & 84.069337 & 34.083562  & 12.10       &   0.17        &    0.06        & 1.50 $\pm$ 0.31 & 162  $\pm$  6    & NM & M (94) \\   
 87 & 83.996719 & 34.174535  & 10.62       &   0.07        &    0.95        & 1.34 $\pm$ 0.16 & 161  $\pm$  3    & M & M (94) \\   
 91 & 84.040637 & 34.193293  & 10.34       &   0.01        &    -0.50        & 1.25 $\pm$ 0.14 & 165  $\pm$  3    & M & M (94) \\   
 92 & 84.024453 & 34.201113  & 10.93       &   0.03        &    -0.49        & 1.16 $\pm$ 0.18 & 164  $\pm$  4    & M & M (94) \\
\hline                                           
\hline                                           
\end{tabular}\\ 
$^{\dagger}$: Boden (1951)\\
$^*$: Sharma et al. (2006)\\  
$^a$: Johnson H.L., Morgan W.W. (1953)\\
$^{\dagger\dagger}$: Present Work\\
$^{\dagger\dagger\dagger}$: Dias et al. (2006)\\
\end{table}

\begin{table}
\caption{Polarimetric results on 21 stars observed towards Stock 8.}\label{result_stock8}
\begin{tabular}{cccccccccc}\hline \hline
Id.$^{\dagger}$&$\alpha (\degr)$          &$\delta (\degr)$            & $V$ & $ (B-V)$($^{*}$) & $ (U-B)$($^{*}$) & $P_{V}$ &  $\theta_{V}$ & M$_{p}$$^{\dagger\dagger}$ &  M$_{p}$$^{\dagger\dagger\dagger}$ \\ 
             &(J2000)& (J2000)&  (mag)$^{*}$ &  &  & $(\%)$   & $(\degr)$  &  & \\
(1) & (2) & (3) & (4) & (5) & (6) & (7) & (8) & (9) & (10) \\
\hline
 11 & 82.004547 & 34.450581  &  11.50  &   0.13  &  0.17   & 1.57 $\pm$0.14 & 163  $\pm$  2  & NM  & M (66)  \\
 12 & 82.009061 & 34.404500  &  11.76  &  -0.36  &  0.28   & 2.41 $\pm$0.15 & 165  $\pm$  2  & M  & M (66)  \\
 13 & 82.018190 & 34.489546  &  11.71  &  -0.47  &  0.30   & 2.45 $\pm$0.15 & 167  $\pm$  2  & M  & M (64)  \\
 16 & 82.038176 & 34.473935  &  11.89  &  -0.40  &  0.29   & 2.41 $\pm$0.16 & 151  $\pm$  2  & M  & M (65)  \\
 19 & 82.058027 & 34.438897  &  13.01  &   1.22  &  1.40   & 2.17 $\pm$0.28 & 157  $\pm$  4  & NM  & M (77)  \\
141 & 82.082569 & 34.420441  &  12.19  &  -0.24  &  0.51   & 2.60 $\pm$0.19 & 165  $\pm$  2  & M  & M (57)  \\
160 & 82.002940 & 34.493490  &  12.63  &   0.15  &  0.60   & 1.62 $\pm$0.23 & 163  $\pm$  4  & NM  & NM (02)  \\
173 & 82.082563 & 34.455984  &  12.81  &   0.37  &  0.31   & 1.60 $\pm$0.25 & 159  $\pm$  4  & NM  & NM (47)  \\
181 & 82.064303 & 34.423842  &  12.87  &  -0.32  &  0.33   & 2.39 $\pm$0.25 & 158  $\pm$  3  & M  & M (68)  \\
208 & 82.040118 & 34.446178  &  13.18  &  -0.14  &  0.30   & 2.62 $\pm$0.28 & 162  $\pm$  3  & M  & M (74)  \\
211 & 81.996346 & 34.444053  &  13.21  &   0.17  &  0.28   & 1.93 $\pm$0.30 & 159  $\pm$  4  & NM  & NM (19)  \\
218 & 82.004829 & 34.385705  &  13.29  &  -0.02  &  0.34   & 2.61 $\pm$0.31 & 162  $\pm$  3  & M  & NM (43)  \\
231 & 82.028692 & 34.394823  &  13.36  &  -0.14  &  0.31   & 3.03 $\pm$0.33 & 157  $\pm$  3  & M  & M (72)  \\
258 & 81.969263 & 34.389477  &  13.53  &   0.24  &  0.68   & 2.57 $\pm$0.35 & 161  $\pm$  4  & M  & NM (00)  \\
275 & 82.090470 & 34.409075  &  13.69  &  -0.17  &  0.62   & 2.37 $\pm$0.39 & 136  $\pm$  5  & ? &  -     \\
294 & 82.003915 & 34.403745  &  13.81  &   0.38  &  0.46   & 2.72 $\pm$0.40 & 163  $\pm$  4  & M  & M (52)  \\
320 & 81.993454 & 34.470836  &  13.93  &   0.27  &  0.62   & 2.20 $\pm$0.45 & 156  $\pm$  6  & M  & NM (01)  \\
325 & 81.962444 & 34.428255  &  13.96  &  -0.08  &  0.82   & 4.18 $\pm$0.44 & 165  $\pm$  3  & ?  & M (71)  \\
333 & 82.066055 & 34.399693  &  14.00  &   0.32  &  0.59   & 2.40 $\pm$0.45 & 163  $\pm$  5  & M  &  -     \\
365 & 82.001149 & 34.457243  &  14.18  &   0.37  &  0.41   & 2.17 $\pm$0.46 & 165  $\pm$  6  & NM  &  -     \\
425 & 81.991192 & 34.498474  &  14.39  &   0.19  &  0.55   & 2.36 $\pm$0.61 & 158  $\pm$  7  & M  & NM (04)  \\

\hline                                           
\hline                                           
\end{tabular}\\
$^{\dagger}$:  Mayer (1964) \\
$^{*}$: Jose et al. (2008) \\   
$^{\dagger\dagger}$: Present Work\\
$^{\dagger\dagger\dagger}$: Dias et al. (2006)\\
\end{table}


\begin{table}
\caption{The $P_{max}$, $\lambda_{max}$, $\sigma_{1}$ \& $\overline\epsilon$ for the observed 44 stars towards NGC 1893}\label{tab_dust_prop}
\begin{tabular}{cccccccccc} \hline  \hline
Id.$^{\dagger}$ &  $E(B-V)$$^{\ddagger}$ & $\alpha$ ($\degr$)  & $\delta$ ($\degr$) & $P_{max}\pm\epsilon$ & $\lambda_{max}\pm\epsilon$ & $\sigma_{1}$ & $\overline\epsilon$ & $M_P$$^{\dagger\dagger}$ & $M_P$$^{\dagger\dagger\dagger}$   \\ 
          &      (mag)   & (J2000)  & (J2000) & $(\%)$   & $(\mu m)$ &  & & \\     
(1)&(2)& (3)& (4) & (5) & (6) & (7) & (8) & (9) & (10) \\                                                               
\hline \hline
  1   &    0.53   &       80.683340   &  33.440739   &  2.50   $\pm$  0.08   &  0.53   $\pm$  0.04   &  0.74   &  1.31   &  M    & M (78)     \\
  2   &    0.54   &       80.686558   &  33.443501   &  3.11   $\pm$  0.13   &  0.54   $\pm$  0.05   &  0.69   &  1.23   &  M    & M (79)     \\
  3   &    0.45   &       80.706563   &  33.447926   &  2.41   $\pm$  0.08   &  0.58   $\pm$  0.04   &  0.91   &  1.38   &  M    & M (78)     \\
  5   &    0.46   &       80.725969   &  33.445042   &  2.52   $\pm$  0.26   &  0.52   $\pm$  0.10   &  0.59   &  0.47   &  M    & NM (9)     \\
  9   &    0.52   &       80.706864   &  33.425991   &  2.76   $\pm$  0.18   &  0.53   $\pm$  0.07   &  0.89   &  1.25   &  M    & M (75)     \\
 10   &    0.46   &       80.705199   &  33.428371   &  2.61   $\pm$  0.28   &  0.70   $\pm$  0.14   &  1.03   &  0.41   &  M    & NM (0)     \\
 13   &    0.52   &       80.692334   &  33.422447   &  2.80   $\pm$  0.22   &  0.62   $\pm$  0.10   &  0.54   &  0.54   &  M    & M (79)     \\
 15   &    0.48   &       80.696872   &  33.418720   &  2.89   $\pm$  0.25   &  0.57   $\pm$  0.10   &  0.83   &  0.92   &  M    & M (61)     \\
 33   &    0.50   &       80.688066   &  33.406540   &  2.32   $\pm$  0.20   &  0.56   $\pm$  0.10   &  0.62   &  1.27   &  M    & M (71)     \\
 48   &    0.20$^{a}$ &   80.699353   &  33.476135   &  1.76   $\pm$  0.14   &  0.51   $\pm$  0.06   &  1.46   &  0.17   & NM    & M (60)     \\
 59   &    0.47   &       80.741815   &  33.443459   &  2.26   $\pm$  0.13   &  0.60   $\pm$  0.07   &  0.72   &  0.98   &  M    & M (79)     \\
 62   &    0.47   &       80.739711   &  33.433380   &  2.41   $\pm$  0.16   &  0.56   $\pm$  0.07   &  0.82   &  0.65   &  M    & M (74)     \\
 72   &     0.20$^{a}$&   80.699203   &  33.368629   &  1.86   $\pm$  0.40   &  0.87   $\pm$  0.23   &  0.82   &  1.05   & NM    & NM (27)    \\
 73   &    0.27   &       80.688608   &  33.371365   &  1.80   $\pm$  0.11   &  0.53   $\pm$  0.06   &  0.58   &  0.50   & NM    & M (79)     \\
 82   &    0.70   &       80.654111   &  33.386971   &  3.32   $\pm$  0.26   &  0.63   $\pm$  0.10   &  0.62   &  0.87   & M    & M (77)     \\
 92   &    0.44   &       80.668795   &  33.491486   &  1.48   $\pm$  0.15   &  0.59   $\pm$  0.12   &  0.31   &  0.58   &  NM    & M (74)     \\
 98   &    0.68   &       80.722431   &  33.509064   &  3.55   $\pm$  0.35   &  0.50   $\pm$  0.09   &  0.56   &  0.49   &  M    & NM (3)     \\
106   &    0.52   &       80.751932   &  33.496628   &  2.87   $\pm$  0.16   &  0.53   $\pm$  0.06   &  0.90   &  0.49   &  M    & M (74)     \\
108   &    0.46   &       80.765402   &  33.487503   &  3.13   $\pm$  0.34   &  0.52   $\pm$  0.10   &  0.97   &  0.56   &  ?    & NM (45)     \\
115   &    0.44   &       80.767404   &  33.470600   &  2.90   $\pm$  0.24   &  0.56   $\pm$  0.09   &  0.43   &  0.38   &  M    &   -     \\
123   &    0.45   &       80.782318   &  33.467247   &  3.77   $\pm$  0.27   &  0.49   $\pm$  0.06   &  1.56   &  0.65   &  ?    &   -     \\
134   &    0.48   &       80.744271   &  33.400803   &  3.19   $\pm$  0.21   &  0.57   $\pm$  0.08   &  0.92   &  1.49   &  M    & M (79)     \\
139   &    0.40   &       80.723816   &  33.392075   &  2.77   $\pm$  0.19   &  0.65   $\pm$  0.08   &  0.57   &  0.10   &  M    & M (78)     \\
140   &    0.41   &       80.719063   &  33.386875   &  2.72   $\pm$  0.19   &  0.57   $\pm$  0.09   &  0.75   &  0.67   &  M    & M (72)     \\
141   &    0.42   &       80.717616   &  33.384277   &  2.60   $\pm$  0.13   &  0.57   $\pm$  0.06   &  0.67   &  0.80   &  M    & M (78)     \\
149   &    0.77   &       80.665351   &  33.371723   &  2.81   $\pm$  0.07   &  0.56   $\pm$  0.03   &  1.35   &  2.43   &  M    & M (80)     \\
168   &    0.53   &       80.643091   &  33.489128   &  2.34   $\pm$  0.20   &  0.60   $\pm$  0.10   &  0.66   &  0.81   &  M    & M (76)     \\
196   &   0.83$^*$&       80.788871   &  33.500652   &  2.26   $\pm$  0.15   &  0.61   $\pm$  0.08   &  0.73   &  1.21   &  ?    & M (79)     \\
197   &    0.20$^{a}$&    80.781174   &  33.497807   &  2.19   $\pm$  0.13   &  0.57   $\pm$  0.06   &  0.96   &  0.60   &  NM    & NM (0)     \\
199   &    0.73   &       80.781643   &  33.477089   &  2.82   $\pm$  0.24   &  0.54   $\pm$  0.08   &  2.19   &  0.98   &  ?    &   -     \\
209   &   0.25$^*$&       80.823050   &  33.436195   &  2.43   $\pm$  0.30   &  0.41   $\pm$  0.06   &  1.50   &  0.46   & NM    & M (80)     \\
213   &   0.53$^*$&       80.798357   &  33.402302   &  3.41   $\pm$  0.46   &  0.51   $\pm$  0.11   &  0.84   &  1.45   &  ?    & NM (22)      \\
215   &    0.47   &      80.789509   &  33.406975   &  3.23   $\pm$  0.67   &  0.41   $\pm$  0.10   &  0.40   &  0.97   &  M    & M (81)     \\
219   &    0.48   &      80.782304   &  33.385307   &  3.81   $\pm$  0.36   &  0.52   $\pm$  0.08   &  1.11   &  0.78   &  M    & M (77)     \\
222   &    0.39   &      80.770831   &  33.389034   &  2.54   $\pm$  0.33   &  0.57   $\pm$  0.15   &  0.21   &  1.16   &  M    & M (78)     \\
228   &    0.40   &      80.741278   &  33.368721   &  2.81   $\pm$  0.21   &  0.62   $\pm$  0.09   &  1.17   &  0.66   &  M    & M (77)     \\
301   &   0.39$^*$&      80.831142   &  33.452225   &  3.21   $\pm$  0.07   &  0.55   $\pm$  0.03   &  1.62   &  0.69   &  M    & M (78)     \\
307   &   0.42$^*$&      80.854608   &  33.435932   &  3.44   $\pm$  0.10   &  0.57   $\pm$  0.04   &  0.28   &  0.28   &  M    & M (79)     \\
310   &   0.20$^{a}$&    80.847118   &  33.419819   &  2.15   $\pm$  0.30   &  0.45   $\pm$  0.07   &  0.49   &  0.74   & NM    & M (84)     \\
320   &   0.46$^*$&      80.802256   &  33.366192   &  2.81   $\pm$  0.17   &  0.55   $\pm$  0.08   &  0.25   &  0.56   &  M    & M (65)     \\
327   &   0.20$^*$&      80.680573   &  33.288097   &  1.16   $\pm$  0.14   &  0.50   $\pm$  0.13   &  0.68   &  0.60   & NM    & M (75)     \\
329   &   0.19$^*$&      80.656141   &  33.294849   &  1.15   $\pm$  0.20   &  0.67   $\pm$  0.20   &  1.22   &  0.83   & NM    & M (76)     \\
355   &   0.20$^{a}$&    80.899315   &  33.423325   &  2.05   $\pm$  0.35   &  0.68   $\pm$  0.20   &  0.09   &  0.53   & NM    & NM (48)     \\
356   &   0.20$^{a}$&    80.853457   &  33.408504   &  2.59   $\pm$  0.36   &  0.57   $\pm$  0.16   &  1.18   &  0.75   & NM    &  -      \\

 \hline \hline
\end{tabular}\\
$^{\dagger}$: Cuffey and Shapley (1937)\\
$^{\ddagger}$: To estimate $E(B-V)$ values, UBV photometric data has been taken from Sharma et al. (2007)\\
$^{*}$: To estimate $E(B-V)$ values, UBV photometric data has been taken from Massey et al. (1995)\\
$^{a}$: The foreground stars with $E(B-V)$=0.20 mag from ZAMS fitting \\
$^{\dagger\dagger}$: Present Work\\
$^{\dagger\dagger\dagger}$: Dias et al. (2006)\\

\end{table}

\begin{table*}
\centering
\caption{Mean values of $P_{V}$ and $\theta_{V}$ for members and non-members.}\label{Mean_M_NM}
\begin{tabular}{lccc}\hline \hline
Cluster Id &  $P_{V}\pm\sigma (\%)$ &  $\theta_{V}\pm\sigma (\degr)$  & No of stars  \\
\hline
           &   Members              &                                  &  \\
\hline
NGC 2281   &  0.9$\pm$0.2 & 17$\pm$6  & 11  \\
NGC 1664   &  1.4$\pm$0.2 & 172$\pm$4 & 18 \\
NGC 1960   &  1.2$\pm$0.1 & 160$\pm$3 & 12 \\
Stock 8    &  2.5$\pm$0.2 & 161$\pm$4 & 13 \\
NGC 1893   &  2.8$\pm$0.4 & 160$\pm$4 & 28 \\
\hline
           &  Non-Members           &        &  \\
\hline
NGC 2281   &  0.9$\pm$0.3 & 14$\pm$8  & 3  \\
NGC 1664   &  2.5$\pm$0.9 & 171$\pm$14 & 9 \\
NGC 1960   &  1.2$\pm$0.4 & 155$\pm$7 & 3 \\
Stock 8    &  1.8$\pm$0.3 & 161$\pm$3 & 6 \\
NGC 1893   &  1.6$\pm$0.4 & 166$\pm$4 & 11 \\
\hline
\hline
\end{tabular}
\end{table*}

\label{lastpage}
\end{document}